\documentclass[arXiv,
]{actapoly}

\usepackage[utf8]{inputenc}

\begin{document}

\title[Beta Functions in Six-Derivative Quantum Gravity]
{How to understand the Structure of Beta Functions in Six-derivative Quantum Gravity?}

\correspondingauthor[L. Rachwa\l{}]{Les\l{}aw Rachwa\l{}}{my,second}{grzerach@gmail.com}

\institution{my}{Departamento de F\'{i}sica---Instituto de Ci\^{e}ncias Exatas,
Universidade Federal de Juiz de Fora,\\ 33036-900, Juiz de Fora, MG, Brazil}
\institution{second}{speaker at AAMP XVIII, online session}

\begin{abstract}
We extensively motivate the studies of higher-derivative gravities, and in particular we emphasize which
new quantum features theories with six derivatives in their definitions  possess. Next, we discuss the mathematical structure
of the exact on the full quantum level  beta functions obtained previously for three couplings in front of generally covariant terms
with four derivatives (Weyl tensor squared, Ricci scalar squared and the Gauss-Bonnet scalar) in minimal
six-derivative quantum gravity in $d=4$ spacetime dimensions. The fundamental role here is played by the
ratio $x$ of the coupling in front of the term with Weyl tensors to the coupling in front of the term
with Ricci scalars in the original action. We draw a relation between the polynomial dependence on $x$ and the absence/presence of
enhanced conformal symmetry and renormalizability in the models where formally $x\to+\infty$ in the case of four- and six-derivative
theories respectively.
\end{abstract}


\maketitle

\section{Introduction and motivation}
\label{s1}

\indent Six-derivative Quantum Gravity (QG) is a model of quantum dynamics of relativistic gravitational field with higher derivatives. It is a special case of general higher-derivative (HD) models which are very particular modifications of Einsteinian gravitational theory. The last one is based on the theory with up to two derivatives (an addition of the cosmological constant term brings terms with no derivatives on the metric field at all) and it is simply based on the action composed of the Ricci curvature scalar $R$ understood as the function of the spacetime metric. In this setup, we consider that gravitational field is completely described by the symmetric tensor field $g_{\mu\nu}$ being the metric tensor of the pseudo-Riemannian smooth differential manifold of a physical spacetime. In Einstein's theory the scalar $R$ contains precisely two ordinary (partial) derivatives of the metric. The action obtained by integrating over the spacetime volume the densitized Lagrangian $\sqrt{|g|}R$ we call as Einstein-Hilbert action. The QG models based on it were originally studied in \cite{DeWitt:1967yk,DeWitt:1967ub,DeWitt:1967uc}. Below we consider modifications of two-derivative gravitational theory, where the number of derivatives on the metric is higher than just two.

It must be remarked, however, that the kinematical framework of general relativity (GR) (like metric structure of the spacetime manifold, the form of Christoffel coefficients, the motion of probe particles, or geodesic and fluid dynamics equations) remains intact for these modifications. Therefore these higher-derivative (HD) models of gravitational field are still consistent with the physical basis of GR, the only difference is that their dynamics --  the dynamics of the gravitational field -- is described by classical equations of motion with higher-derivative character. Thence these modifications of standard Einsteinian gravitational theory are still in the set of generally relativistic models of the dynamics of the gravitational field. They could be considered both on the classical and quantum levels with the benefits of getting new and deeper insights in the theory of relativistic gravitational field. Our framework on the classical level can be summarized by saying that we work within the set of metric theories of gravity, where the metric and only the metric tensor characterizes fully the configurations of the gravitational fields which are here represented by pseudo-Riemannian differential manifolds of relativistic four-dimensional continuous spacetimes. Therefore, in this work we neglect other classical modifications of GR, like by adding torsion, non-metricity, other geometric elements or other scalars, vectors or tensor fields. This choice of the dynamical variables for the relativistic gravitational field bears impact both on the classical dynamics as well as on the quantum theory.

Theories with higher derivatives come naturally both with advantages and with some theoretical problems. This happens already on the classical level when they supposed to describe the modified dynamics of the gravitational field (metric field $g_{\mu\nu}(x)$ living on the spacetime manifold). These successes and problems get amplified even more on the quantum level. The pros for HD theories give strong motivations why to consider seriously these modifications of Einstein's gravitation. We will briefly discuss various possibilities of how to resolve the problems of higher-derivative field dynamics in one of the last sections of this contribution, while here we will consider more the motivations.

On the classical level, the set of HD gravitational theories can be viewed as one of the many possible modifications of two-derivative gravitational theory. It is true that now observations, mainly in cosmology and on the intergalactic scales, point to some possible failures of Einsteinian theory of gravity or to our lack in understanding the proper nature of the sources of gravity in these respective situations. There are various views possible on this situation and its explanations by gravitational theories. In the first view, researchers say that Einstein-Hilbert theory is still fine, but we need to add locally new exotic (meaning coming with some non-standard properties) matter source. Since we do not know what these sources for energy-momentum tensor (EMT) of matter are built out of (for example -- from which quantum fields of particle physics as understood nowadays), we call the missing sources as dark energy and dark matter respectively. Contrary to this approach, in the other, the gravitational source is standard, that is we describe what we really see in the galaxies and in the universe, without any ``dark'' components, but the gravitational theory should be modified. In this second path, the internal dynamics of the gravitational field is changed and that is why it reacts differently to the same classical visible EMT source of standard matter. One of the promising options is to add higher derivatives of the metric on the classical level, but in such a way to still preserve the local Lorentz symmetry of the dynamics that is to be safe with respect to the general covariance. Hence all HD terms in the action of the theory must come from generally densitized scalars which are HD analogs of the Ricci scalar. They can be in full generality built as contractions of the metric tensors (both covariant $g_{\mu\nu}$ and contravariant $g^{\mu\nu}$),  Riemann curvature tensor $R_{\mu\nu\!\rho\sigma}$ and also of covariant derivatives $\nabla_\mu$ acting on these Riemann tensors\footnote{We do not need to consider covariant derivatives on the metric tensor because of the metricity condition, $\nabla_\mu g_{\nu\!\rho}=0$.}. Initially this may look as presumably unnecessary complication since classical equations of motion (EOM) with higher derivatives of the gravitational field are even more complicated than already a coupled system of non-linear partial differential equations for the components of the metric tensor field  in Einstein's gravity. However, on the cosmological and galactic scales some gravitational  models with higher derivatives give successes in explaining: the problem of dark matter halos, flat galactic rotation curves, cosmological dark energy (late-time exponential expansion of the universe) and also primordial inflation without a necessity of having the actual inflaton field. These are amongst all of observational pieces of evidence that can be taken for HD models.

Since our work is theoretical we provide below some conceptual and consistency arguments for HD gravities. First, still on the classical level, within the class of higher-derivative gravitational theories, there are models that are the first, which besides relativistic symmetries, enjoy also invariance under conformal symmetry  understood in the  GR framework. Properly this is called as Weyl symmetry of the rescaling of the covariant metric tensor, according to the law: $g_{\mu\nu} \to \Omega^2 g_{\mu\nu}$  with $\Omega=\Omega(x)$ being an arbitrary scalar parameter of these transformations. To understand better this fact, one may first recall that the metric tensor $g_{\mu\nu}$ is taken here as a dimensionless quantity and all energy dimensions are brought only by partial derivatives acting on it. Next, the prerequisite for full conformal symmetry is scale-invariance of the classical action, so the absence of any dimensionful parameter in the definition of the theory. From these facts, one derives that in four spacetime dimensions ($d=4$) the gravitational conformal models must possess terms with precisely four derivatives acting on the metric. In general, in $d$ dimensions, for conformal gravitational theory the classical action  must be precisely with $d$ derivatives on the metric. (One sees due to the requirement of general covariance that this consideration of conformal gravitational theories makes sense only in even dimensions $d$ of spacetime.) Another interesting observation, is that the gravitational theory with Einstein-Hilbert action is classically conformally invariant only in two-dimensional framework. For $4$-dimensional scale-invariant gravitational theory one must use a combination of the squares of the Riemann tensors and various contractions thereof. (The term $\square R$ is trivially a total derivative term, so cannot be used.) Therefore for $d$-dimensional conformal gravitational theories ($d>2$) we inevitably must consider HD metric theories. The conformally invariant gravitational dynamics is very special both on the classical level and also on the quantum level as we will see in the next sections.

The main arguments for higher-derivative gravitational theories in dimensions $d>2$ come instead from quantum considerations. After all, it is not so surprising that it is the quantum coupling between quantum field theory (QFT) of matter fields and quantum (or semi-classical) gravity or self-interactions within pure quantum gravity that dictates what should be a consistent quantum theory of gravitational interactions. Our initial guess (actually Einstein's one) might not be the best one when quantum effects are taken fully into account. Since it is the classical theory that is emergent from the more fundamental quantum one working not only in the microworld, but at all energy scales (equivalent to various distances), then the underlying fundamental quantum theory must necessarily be mathematically consistent, while some different classical theories may not possess the same strong feature. Already here we turn the reader's attention to the fact that the purely mathematical requirement of the consistency on the quantum level of gravitational self-interactions is very strongly constraining the possibilities for quantum gravitational theories. It is more constraining than it was originally thought of. Moreover, not all macroscopic, so long wavelength limit, classical theories are with these quantum correspondence features, only those which emerge as classical limits of consistent quantum gravity theories. Following this path, at the end, we must also correct our classical gravitational theory, and likely it will not be Einsteinian gravity any more.

From a different side, we know that matter fields are quantum, they interact and they are energetic, so they are ``charged'' under gravitation since energy-momentum content is what the gravity couples to. If we did not know nothing about gravity, then we could discover something about it from quantum considerations of gravitationally ``charged'' matter fields and their mutual interactions consistent with quantum mechanics. In this way we could make gravity dynamical and quantum with a proper form of graviton's propagation. Actually, it is the quantum consideration that makes the gauge bosons mediating the interactions between quantum charged particles dynamical. These gauge bosons are emanations or quantum realizations of classical dynamical gauge fields that must be introduced in the classical dynamics of matter fields or particles for the overall consistency. Below we will present a few detailed arguments why we need HD gravities in $d>2$ giving rise to dynamical gravitational fields with HD form of the graviton's propagators in the quantum domain. They are all related and in a sense all touch upon the issue of coupling of a potential unknown dynamical quantum gravity theory to some energetic quantum matter fields moving under the influence of classical initially non-dynamical gravitational background field.  (The background gravitational field does not have to be static, stationary or completely time-independent, what  we only require here is that it is not a dynamical one.) These last classical fields can be understood as frozen expectation values of some dynamical quantum gravitational fields. As one can imagine for this process of quantum balancing of interactions the issue of back-reaction of quantum matter fields on the classical non-dynamical geometry is essential.

Firstly, we recall the argument of DeWitt and Utiyama \cite{Utiyama:1962sn}. Due to quantum matter loops some UV divergences in the gravitational sector are generated. This is so even if the original matter theory is with two-derivative actions (like for example standard model of particle physics). The reasons for these divergences are pictorially Feynman diagrams with quantum matter fields running in the perturbative loops, while the graviton lines are only external lines of the diagrams since they constitute classical backgrounds. In such a way we generate the dynamics to the gravitational field due to quantum matter interactions with gravity, so due to the back-reaction phenomena. If the latter was neglected we would have only the impact of classical gravitational field on the motion and interactions of quantum matter particles.
We can be very concrete here, namely for example in $d=4$ spacetime dimensions, the dynamical action that is generated for gravity takes the form
\begin{equation}
S_{\rm div} = \!\int\! d^4x \sqrt{|g|}\left(\alpha_C C^2+\alpha_R R^2\right),
\label{sdiv}
\end{equation}
so we see that counterterms of the GR-covariant form of $C^2$ and $R^2$ are being generated. (In the equation above, the $R^2$ and $C^2$ terms denote respectively the square of the Ricci scalar and of the Weyl tensor, where the indices are contracted in the natural order, i.e. $C^2=C_{\mu\nu\!\rho\sigma}C^{\mu\nu\!\rho\sigma}$. Collectively, we will denote these curvatures as ${\cal R}^2$, so ${\cal R}^2=R^2,C^2$.) This is true no matter what was our intention of what was the dynamical theory of the gravitational field. We might have thought that this was described by the standard two-derivative Einstein-Hilbert action, but still the above results persist. One notices that in these two counterterms $C^2$ and $R^2$ one has four derivatives acting on a metric tensor, so these are theories of a general higher-derivative type, differently from originally intended E-H gravitational theory whose action is just based on the  Ricci scalar $R$. These $C^2$ and $R^2$ terms appear in the divergent part of the dynamically induced action for the gravitational fields. We must be able to absorb these divergences to have a consistent quantum theory of the gravitational field coupled to the quantum matter fields present here on such curved (gravitational) backgrounds \cite{Birrell:1982ix}. This implies that in the dynamics of the gravitational field we must have exactly these terms with higher derivatives as in (\ref{sdiv}). Finally, we can even abstract and forget about matter species and consider only pure gravitational quantum theory. The consistency of self-interactions there on the quantum level puts the same restriction on the form of the action of the theory. In such a situation in the language of Feynman diagrams, one considers also loops with quantum gravitons running inside. These graphs induce the same form of UV divergences as in (\ref{sdiv}). Then in such a model, we must still consider the dynamics of the quantum gravitational field with higher derivatives. Hence, from quantum considerations higher derivatives are inevitable.

We also remark here that  in the special case, where the matter theory is classically conformally invariant with respect to classical gravitational background field (the examples are: massless fermion, massless Klein-Gordon scalar field conformally coupled to the geometry, electrodynamic field or non-Abelian Yang-Mills field in $d=4$), then only the conformally covariant counterterm $C^2$  is generated, while the coefficient $\alpha_R=0$ in (\ref{sdiv}). This is due to the fact that the quantization procedure preserves conformal symmetries of the original classical theory coupled to the non-trivial spacetime background. Such argument can be called as a conformal version of the original DeWitt-Utiyama argument. Then the $R^2$ counterterm is not needed but still the action of a quantum consistent coupled conformal system requires the higher-derivative dynamics in the gravitational sector \cite{Buchbinder:1992rb}.  Here this is clearly the gravitational dynamics only in the spin-$2$ sector of metric fluctuations, which is contained entirely in the (conformal) $C^2$ sector of the generic four-derivative theory presented in (\ref{sdiv}).

An intriguing possibility for having higher derivatives in the gravitational action was first considered by Stelle in \cite{Stelle:1976gc} and some exact classical solutions of such a theory were analyzed in \cite{Stelle:1977ry,Lu:2015psa,Lu:2015cqa}. In $d=4$ spacetime dimensions, the minimal number of derivatives is exactly four, the same as the number of dimensions \cite{Modesto:2011kw}. This reasoning coincides  with the one presented earlier that we need to have in even number of dimensions $d$, precisely $d$ derivatives in the gravitational action to have first scale-invariant model of gravitational dynamics (later possible to be promoted to enjoy also the full conformal invariance). However, as proven by Asorey, Lopez and Shapiro in \cite{Asorey:1996hz}, there are also possible theories with even  higher number of derivatives, and they still have good properties on the quantum level and when coupled to quantum matter fields. Similarly, in the literature there are various known motivations for conformal gravity in $d=4$ spacetime dimensions, one can consult representatives in \cite{Mannheim:2011ds,Mannheim:2016gwb}.

Secondly, we emphasize that to have a minimal (in a sense with the smallest number of derivatives) perturbatively renormalizable model of QG in dimensions $d$, one also has to consider actions with precisely $d$ derivatives. The actions with smaller number are not scale-invariant and have problems on the quantum level to control all perturbative UV divergences, and not all of them are absorbable in the counterterms coming from the original classical actions of the theories -- such models with less than $d$ derivatives are not multiplicatively renormalizable. The first case for renormalizability is when the action contains all generic terms with arbitrary coupling coefficients with $d$ (partial) derivatives on the metric for $d$ dimensions. The special cases when some coefficients and some coupling parameters vanish may lead to restricted situations in which full renormalizability is not realized. We discuss such special limiting cases in further sections of this paper. The argument with the first renormalizable theory is a very similar in type to the quantum induced action from matter fields, but this time the particles which run in the perturbative loops of Feynman diagrams are quantum gravitons themselves. So this argument about renormalizability applies to pure quantum gravity cases. Unfortunately, the original Einstein-Hilbert action for QG model is not renormalizable (at least not perturbatively) in $d=4$ dimensions \cite{tHooft:1974toh,Deser:1974cz,Goroff:1985sz,Goroff:1985th}. The problems show up when one goes off-shell, couples some matter, or goes to the two-loop order, while at the first loop order with pure E-H gravitational action on-shell all UV divergences could be successfully absorbed \cite{tHooft:1974toh} on Einstein vacuum backgrounds (so on Ricci-flat configurations). Actually, in such a case in vacuum configurations the theory at the one-loop level is completely UV-finite.

There are also other ways how one can on the quantum level induce the higher-derivative terms in the gravitational actions, although these further arguments are all related to the original one from DeWitt and Utiyama. One can, for example, consider integrating out completely quantum matter species on the level of functional integral which represents all accessible information of the quantum theory. In the situation, when these matter species are coupled to some background gravitational field, then the resulting partition function $Z$ is a functional of the background gravitational field. Not surprisingly, this functional is of the higher-derivative nature in terms of number of derivatives of the fundamental metric field, if we work in the dimension $d>2$. This reasoning was for example popularized by 't Hooft \cite{tHooft:2014swy,tHooft:2015vaz,tHooft:2016uxd}, especially since in $d=4$ it can give rise to another motivations for conformal gravity as a quantum consistent model of conformal and gravitational interactions, when massless fields are integrated out in the path integral.

In this way we can discover the quantum consistent dynamics of the gravitational field even if we did not know that such quantum fields mediating gravitational interactions between particles existed in the first place. The graviton becomes a propagating particle and with higher-derivative form of the propagator, which translates in momentum space to the enhanced suppression of the fall-off of the propagator for large momenta in the UV regime. This is due to the additional higher powers of propagating  momentum in the perturbative expression for the graviton's propagator. This enhanced UV decaying form of the propagator is what makes the UV divergences under perturbative control and what makes the theory at the end renormalizable. Besides a few (finite number of) controlled UV divergences the theory is convergent and gives finite perturbative answers to many questions one can pose about the quantum dynamics of the gravitational field, also in models coupled consistently to quantum matter fields.

Another way is to consider the theory of Einsteinian gravity and corrections to it coming  from higher dimensional theories. One should already understood from the discussion above, that E-H action is a good quantum action for the QG model only in the special $2$-dimensional case. There in $d=2$  QG is very special renormalizable and finite theory, but without dynamical content resembling anything what is known from four dimensions (like for example the existence of gravitational waves, graviton spin-2 particles, etc.). This is again due to infinite power of conformal symmetry in $d=2$ case. Instead, if one considers higher dimensions like $6$, $8$, etc. and then compactifies them to common $4$-dimensional case, one finds that even if in the higher dimensions one had to deal with the two-derivative theory based on the Einstein-Hilbert action, then in the reduced case in four dimensions, one again finds effective (dimensionally reduced) action with four derivatives. These types of arguments were recently invoked by Maldacena \cite{Maldacena:2011mk} in order to study higher-derivative (and conformal) gravities from the point of view of higher dimensions, when the process of integration out of quantum modes already took place and one derives a new dynamics for the gravitational field based on some compactification arguments.

All this above shows that many arguments from even various different directions lead to the studies of higher-derivative gravitational theories in dimensions of spacetime $d>2$. Therefore, it is very natural to quantize such four-derivative theories (like it was first done by Stelle) and treat them as a starting point for discussion of QG models in $d=4$ case. At the end, one can also come back and try to solve for exact solutions of these higher-derivative gravitational theories on the classical level, although due to increased level of non-linearities this is a very difficult task \cite{Li:2015bqa}.

Yet another argument is based on apparent similarity and symmetry seen in the action of quadratic gravity and action for a general Yang-Mills theory. Both these actions are quadratic in the corresponding field strengths (or curvatures). They are curvatures respectively in the external spacetime for the gravitational field and in the internal space for gauge degrees of freedom. The Einstein-Hilbert action is therefore \emph{not similar} to the $F^2$ action of Yang-Mills theory and the system of Einstein-Maxwell or Einstein-Yang-Mills theory does not look symmetric since the number of curvatures in two sectors is not properly balanced. Of course, this lack of balance is later even amplified to the problematic level by quantum corrections and the presence of unbalanced UV divergences (non-renormalizability!). Still, already on the classical level, one sees some dichotomy, especially when one tries to define a common total covariant derivative $D_\mu$ (covariant both with respect to Yang-Mills internal group $G$ and with respect to gravitational field). For such an object, one can define the curvature ${\cal F}_{\mu\nu}$ that is decomposed in its respective sectors into the gauge field strength $F_{\mu\nu}$ and the Riemann gravitational tensor $R_{\mu\nu\!\rho\sigma}$. But the most natural thing to do here is to consider symmetric action constructed with such a total curvature of the derivative $D_\mu$ and then the generalized ${\cal F}^2 $  is the first consistent option to include both dynamics of the non-Abelian gauge field and also of the gravitational field. As we have seen this choice is also stable quantum-mechanically \cite{Buchbinder:1989jd} since there are no corrections that would destabilize it and the only quantum corrections present they support this ${\cal F}^2$ structure of the theory, even if this was not there from the beginning. We emphasize that this was inevitably the higher-derivative structure for the dynamics of the quantum relativistic gravitational field studied here.

\subsection{Motivations for and introduction to six-derivative gravitational theories}
\label{s1.1}

Now, we would like to summarize here on what is the general procedure to define the gravitational theory, both on the classical as well as on the quantum level. First, we decide what our theory is of -- which fields are dynamical there. In our case these are gravitational fields entirely characterized by the metric tensor of gravitational spacetime. Secondly, we specify the set of symmetries (invariance group) of our theory. Again, in our setup these are, in general, invariances under general coordinate transformations also known as diffeomorphism symmetries of gravitational theories. In this sense, we also restrict the set of possible theories from general models considered in the gauge treatment of gravity, when the translation group or full Poincar\'{e} groups are gauged. Then finally following Landau we define the theory by specifying its dynamical action functional. In our case for a classical level, this is a GR-invariant scalar obtained by integrating some GR-densitized scalar Lagrangian over the full 4-dimensional continuum (spacetime). As emphasized above, for theoretical consistency, we must use Lagrangians (actions) which contain higher (partial) derivatives of the metric tensor, when the Lagrangian is completely expanded to a form where ordinary derivatives act on the metric tensors (contracted in various combinations). Specifying  now, to the case motivated above, we shall use and study below the theories defined by classical action functionals which contain precisely six derivatives of the metric tensor field.

In order to define the theory on the quantum level, we use the standard functional integral representation of the partition function (also known as the vacuum transition amplitude) of the quantum theory. That is we construct, having the classical action functional $S_{\rm HD}$, being the functional of the classical metric field $S_{\rm HD}[g_{\mu\nu}]$, the following object
\begin{equation}
Z=\!\int\! {\cal D} g_{\mu\nu} \exp(i S_{\rm HD}),
\end{equation}
where in the functional integral above we must be more careful than just on the formal level in defining properly the integration measure ${\cal D} g_{\mu\nu}$. For example, we should sum over all backgrounds and also over all topologies of the classical background gravitational field. One can hope that it is also possible to classify in four dimensions all gravitational configurations (all gravitational pseudo-Riemannian manifolds) over which we should integrate above. The functional integral, if properly defined, is the basis for quantum theory. One can even promote the point of view that by giving the functional $Z$ one \emph{defines} the quantum theory even without reference to any classical action $S$. However, it is difficult a priori to propose generating functionals $Z$, which are consistent with all symmetries of the theory (especially gauge invariances) and such that they possess sensible macroscopic (classical) limits. For practical purposes of evaluating various correlation functions between quantum fields and their fluctuations, one  modifies this functional $Z$ by adding a coupling of the quantum field (here this role of the integration variable is played by $g_{\mu\nu}$) to the classical external current $J$. And also for other theoretical reasons, one can compute this functional in background field method, where the functional integration is over fluctuation fields, while the classical action functional is decomposed into background and parts quadratic, cubic and of higher order in quantum fluctuation fields. For this one defines that the full metric is decomposed as follows, $g_{\mu\nu} = \bar{g}_{\mu\nu} + h_{\mu\nu}$, where the background classical  metric is denoted by $\bar{g}_{\mu\nu}$ and metric perturbations by $h_{\mu\nu}$.
By computing variational derivatives of the partition function $Z[J]$  with respect to the classical current $J$ one gets higher $n$-point functions with the accuracy of the full quantum level. One can compute them both perturbatively (in loop expansion) or non-perturbatively, and also on trivial backgrounds or in background field method. Finally, for spacetimes which asymptotically reach Riemann-flatness, from on-shell quantum Green functions dressed by wave functions of external classical states, one derives quantum matrix elements of scattering processes. Only in such conditions one can define general scattering problem in quantum gravitational theory.

In this article, we want to analyze the quantum gravitational model with six derivatives in the action.
That the theory is with six derivatives can be seen, because of two related reasons. First, one can
derive the classical equations of motion based on such an action. Then one will see that the number of
partial derivatives acting on a metric tensor in a general term of such tensor of equations of motion is
at most 6 in our model. Or similarly, one can compute the tree-level graviton's propagator for example
around flat Minkowski background. And then one notices that some components of this propagator are
suppressed in the UV regime by the power $k^6$ in Fourier space, when $k$ is the propagating momentum of
the quantum mode. Actually, for this last check one does not even have to invert and compute the
propagator, one can perform a very much the same analysis on the level of the kinetic operator between
gravitational fluctuations around some background (of course the flat background is here the easiest
one). Later in the main text of this article, we discuss how to overcome the problems in defining the
propagator in some special cases, but the situation with the terms of the kinetic operator is almost
always well-defined and one can read the six-derivative character of the theory easily from there.

We have seen in the previous section that the four-derivative gravitational theories in $d=4$ spacetime dimensions are scale-invariant (can be conformally invariant) on the classical level and that they are also first minimal renormalizable models of dynamical QG. This last assertion is proved by the power counting analysis. We will show below that it is possible to further extend the theory in such a way that the control over divergences is strengthened even more and this is again based on the analysis of the superficial degrees of divergences of any graph and also on the energy dimensionality arguments. In this way we will also explain why we can call generic six-derivative gravitational theories in $d=4$ as perturbatively super-renormalizable theories.

The power counting analysis in the case of four-derivative Stelle gravity as in (\ref{sdiv}) (quadratic gravity of the schematic type ${\cal R}^2$ as in \cite{Stelle:1976gc}) leads to the following equality
\begin{equation}
\Delta+d_\partial=4,
\label{pc1}
\end{equation}
where $\Delta$ is superficial degree of divergence of any Feynman graph $G$, $d_\partial$ is the number of derivatives of the metric on the external lines of the diagram $G$, and for future use we define  $L$ as the number of loop order. For tree-level (classical level) we have $L=0$, while for concreteness we shall assume $L\geqslant1$. This theory is simply renormalizable since the needed GR-covariant  counterterms (to absorb perturbative UV divergences) have the same form as the original action in (\ref{sdiv})\footnote{We remind for completeness that the Gauss-Bonnet scalar term ${\rm GB}=E_4=R_{\mu\nu\!\rho\sigma}^2-4R_{\mu\nu}^2+R^2$ is a topological term, that is its variation in four spacetime dimensions leads to total derivative terms contributing nothing to classical EOM and also to quantum perturbation theory. It may however contribute non-perturbatively when the topology changes are expected. But for the sake of computing UV divergences we might simply neglect the presence of this term both in the original action as well as in the resulting one-loop UV-divergent part of the effective action.}.
 In general local perturbatively  renormalizable HD model of QG in $d=4$, the divergences at any loop order must take the form as in (\ref{sdiv}) with a potential addition of the topological Gauss-Bonnet term.

The change in the formula (\ref{pc1}),  when the six-derivative terms are leading in the UV regime, is as follows
\begin{equation}
\Delta+d_\partial=6-2L.
\label{pc2}
\end{equation}
The above formula can be also rewritten as a useful inequality (bound on the superficial degree $\Delta$):
\begin{equation}
\Delta\leqslant6-2L = 4-2(L-1),
\end{equation} since $d_\partial\geqslant0$.
From this one sees an interesting feature that while in the case of four-derivative Stelle theory the bound was independent on the number of loops $L$, for the case of six derivatives (and higher too) the bound is tighter for higher number of loops. This is the basis for super-renormalizability properties. In particular, in the case of six-derivative theories there  are no any loop divergences at the level of fourth loop, since for $L=4$ we find that $\Delta<0$, so all graphs are UV-convergent. We also emphasize that a super-renormalizable model is still renormalizable, but at the same time it is more special since infinities in the former do not show up at arbitrary loop order $L$, which is instead the case for merely renormalizable models. From the formula (\ref{pc2}) at the $L=3$ loop level the possible UV divergences are only of the form proportional to the cosmological constant $\Lambda$ parameter, so completely without any partial derivatives acting on the metric tensor. Similarly for the case of $L=2$, we have that divergences can be proportional to the $\Lambda $ (with no derivatives) and also to the first power of the Ricci scalar $R$ of the manifold (with two derivatives on the metric, when it is expanded). In what follows we will not concentrate on these types of subleading in the UV divergences and our main attention in this paper will be placed on the four-derivative divergences as present in the action (\ref{sdiv}). Up to the presence of the Gauss-Bonnet term they are the same as induced from quantum matter loops. These types of divergences are only generated at the one-loop level since for them we must have $d_\partial=4$ and $\Delta=0$. The last information signifies that they are universal logarithmic divergences. Their names originate from the fact that they arise when the ultraviolet cutoff $\Lambda_{\rm UV}$ is used to cut the one-loop integrations over momenta of modes running in the loop in the upper limits.

The analysis of power counting implies that the theory has divergences only at the first, second and third loop order and starting from the fourth loop level it is completely UV-finite model of QG. Moreover, based on the above argumentation, the beta functions that we report below (in front of GR-covariant terms with four derivatives in the divergent effective action) receive contributions only at the one-loop level and higher orders (like two- and three-loop) do not have any impact on them. This means that the beta functions that we are interested in and that we computed at the one-loop level are all valid to all loop orders, hence our results for them are truly exact. They do not receive any perturbative contributions from higher loops. For other terms in the divergent action (like $\Lambda$ or $R$) this is not true. The theory is four-loop finite, while the beta functions of $R^2$, $C^2$ and ${\rm GB}$ terms are one-loop exact. All these miracles are only possible to happen in very special super-renormalizable model since we have six derivatives in the gravitational propagator around flat spacetime. This number is bigger than the minimal for a renormalizable and scale-invariant QG theory in $d=4$ spacetime dimensions and this is the origin of the facts above since we have a higher momentum suppression in the graviton's propagator.

According to what we have stated before, we decide to study the quantum theory described by the following classical Lagrangian,
\begin{multline}
{\cal L} = \omega_C C_{\mu\nu\!\rho\sigma} \square C^{\mu\nu\!\rho\sigma} + \omega_R R\square R\\
+\theta_C C^2 + \theta_R R^2 +\theta_{\rm GB} {\rm GB}+ \omega_\kappa R+\omega_\Lambda.
\label{lagr}
\end{multline}
From this Lagrangian we construct the action of our HD quantum gravitational model, here with six derivatives as the leading number of derivatives in the UV regime, by the formula
\begin{equation}
S_{\rm HD} = \!\int\! d^4x\sqrt{|g|}{\cal L}.
\end{equation}
Above by $C_{\mu\nu\!\rho\sigma}$ we denote the Weyl tensor (constructed from the Riemann $R_{\mu\nu\!\rho\sigma}$, Ricci tensor $R_{\mu\nu}$ and Ricci scalar $R$ and with coefficients suitable for $d=4$ case). Moreover, by ${\rm GB}$ we mean the Euler term which gives rise to Euler characteristic of the spacetime after integrating over the whole manifold. Its integrand is given by the term also known as the Gauss-Bonnet term and it has the following expansion in other terms quadratic in the gravitational curvatures,
\begin{equation}
{\rm GB}=E_4=R_{\mu\nu\!\rho\sigma}^2-4R_{\mu\nu}^2+R^2.
\label{formulaGB}
\end{equation}
Similarly, we can write for the ``square''  of the Weyl tensor in $d=4$
\begin{equation}
C^2=C_{\mu\nu\!\rho\sigma}^2=C_{\mu\nu\!\rho\sigma}C^{\mu\nu\!\rho\sigma}=R_{\mu\nu\!\rho\sigma}^2-2R_{\mu\nu}^2+\frac13R^2.
\end{equation}
Finally, to denote the box operator we use the symbol $\square$ with the definition $\square=g^{\mu\nu}\nabla_\mu\nabla_\nu$, which is a GR-covariant analogue of the d'Alembertian operator $\partial^2$ known from the flat spacetime.

It is important to emphasize here that the  Lagrangian (\ref{lagr}) describes the most general
six-derivative theory describing the propagation of gravitational fluctuations on flat spacetime. For
this purpose it is important to include  all terms that are quadratic in gravitational curvature. As it
is obvious from the construction of the Lagrangian in (\ref{lagr}) for six-derivative model we have to
include terms which are quadratic in the Weyl tensor or Ricci scalar and they contain precisely one power
of the covariant box operator $\square$ (which is constructed using the
GR-covariant derivative $\nabla_\mu$). These two terms  exhaust all other possibilities  since other terms
which are quadratic and contain two covariant derivatives can be reduced to the two above exploiting
various symmetry properties of the curvature Riemann tensor as well as cyclicity and Bianchi identities.
Moreover, the basis with Weyl tensors and Ricci scalars is the most convenient when one wants to study
the form of the propagator of graviton around flat spacetime. Other bases are possible as well but then
they distort and entangle various contributions of various terms to these propagators. We also remark that
the addition of the Gauss-Bonnet term is possible here (but it is a total derivative in $d=4$); one could
also add a generalized Gauss-Bonnet term, which is an analogue of the formula in  (\ref{formulaGB}), where the
GR-covariant box operator in the first power is inserted in the middle of each of the tensorial terms there, which are
quadratic in curvatures. Eventually, there is no contribution of the generalized Gauss-Bonnet term in any
dimension to the flat spacetime graviton propagator, so for this purpose we do not need to add such term
to the Lagrangian as it was written in (\ref{lagr}).

In what follows we employ the pseudo-Euclidean notations and by $\sqrt{|g|}$  we will denote the square root of the absolute value   of
the metric determinant (always real in our conventions). The two most subleading terms in the Lagrangian (\ref{lagr}) are with couplings $\omega_\kappa$ and $\omega_\Lambda$ respectively. The first one is related to the Newton gravitational constant $G_N$, while the last one $\omega_\Lambda$ to the value of the physical cosmological constant parameter. The QG model  with the Lagrangian (\ref{lagr}) is definitely the simplest one that describes the most general form of the graviton propagator
around flat spacetime, in four spacetime dimensions and for the theory with six derivatives.

We would like to already emphasize here, that there are two remarkable special limiting cases in the theory (\ref{lagr}). In order to
have a non-degenerate classical action and the well-defined Hessian operator of the second variational derivative, one needs to require that both coefficients of the UV-leading terms, namely $\omega_C$ and  $\omega_R$, should be non-zero. Only in this case the theory is renormalizable, moreover only in this case it also has nice additional features like super-renormalizability and that the fourth and higher perturbative loop contributions are completely finite. We want to say that the quantum
calculations reported in the next section correspond only to this kind of well balanced model with both Weyl tensor and Ricci scalar squared terms and one power of the GR-covariant box operator inserted in the middle. (This is in order to have a six-derivative action, but also with terms that are precisely quadratic in gravitational curvatures.) In principle, there exist also models with non-balanced situations and dichotomy between different sectors of fluctuations. For example, in the special case of $\omega_C=0$, $\theta_C\neq0$ and $\omega_R\neq0$, the theory has the propagating spin-two mode with four derivatives and the propagating
spin-zero mode with six derivatives in the perturbative spectrum around flat spacetime. This has to be contrasted with the fact that interaction vertices have always six derivatives in both special and also in generic theories (with $\omega_C\neq0$ and $\omega_R\neq0$).
For another special version of the model, with $\omega_C\neq0$ and $\omega_R=0$, the situation is quite opposite regarding the spectrum, but the negative conclusions are the same. According to the power counting
arguments from \cite{Buchbinder:1992rb,Buchbinder:2021wzv} and also from (\ref{pc1}) in both special cases the theories are unfortunately non-renormalizable. (We also discuss in greater details the power counting for these two special limiting models in section \ref{s4.4}.) Hence one should be very careful in performing computations in such cases and in trusting the results of limits there. These cases will be analyzed in more details in the next sections as it will be revealed that they are crucial for understanding the issue of the structure of perturbative divergences both in the four-derivative as well as also in six-derivative QG models in $d=4$.

The other consequences of the formula for power counting as presented in (\ref{pc2}) is that the subleading in the UV terms of the original action in (\ref{lagr})  do not at all contribute to the four-derivative terms leading in the UV regime of the divergences in (\ref{sdiv}). That is we have that the coefficients $\alpha_C$, $\alpha_R$ and $\alpha_{\rm GB}$  in (\ref{sdiv}) depend only on the ratio of the coefficient in front of the term with Weyl tensors and box inserted in the middle (i.e. $C\square C$) to the coefficient in front of the corresponding term with two Ricci scalars (i.e. $R\square R$), so only on the ratio $\omega_C/\omega_R$ also to be analyzed later at length here. These coefficients of UV divergences $\alpha_C$, $\alpha_R$ and $\alpha_{\rm GB}$ do not depend on $\theta_C$, $\theta_R$, $\theta_{\rm GB}$, $\omega_\kappa$ nor on $\omega_\Lambda$. This is due to the energy dimensionality considerations of other UV-subleading terms in the action in (\ref{lagr}). Only the terms having the same energy dimensionality as the leading in the UV regime (shaping the UV form of the perturbative propagator) may contribute to the leading form  of UV divergences, which in the divergent action (\ref{sdiv})  are represented by dimensionless numbers (in $d=4$)  such as $\alpha_C$, $\alpha_R$  and $\alpha_{\rm GB}$. For example, the terms with coefficients $\theta_C$ or $\theta_R$ have different energy dimensions and cannot appear there. This pertinent observation lets us for our computation to use just the reduced action, where we write only the terms that are important for the UV divergences we want to analyze in this paper. This action takes explicitly the following form
\begin{equation}
S_{\rm HD} = \!\int\!d^4x\sqrt{|g|}\left(\omega_C C_{\mu\nu\!\rho\sigma} \square C^{\mu\nu\!\rho\sigma} + \omega_R R\square R\right).
\label{sred}
\end{equation}

We want to just remark here that the results in the theory with six-derivative gravitational action are
discontinuous to the results one obtains for the similar type of computations in four-derivative Stelle
quadratic QG models, which are usually analyzed in $d=4$ as the first and the most promising models of
higher-derivative QG. This discontinuity is based on the known fact (both for HD gauge and gravitational
theories) that the cases with two and four more derivatives  in the action of respective gauge fields
(metric fields in gravity) than in the minimal renormalizable model are discontinuous and exceptional, while the general formula exists starting
from action with  six derivatives more in its definition (and then this formula could be analytically extended). All
three cases of: first minimal renormalizable theory, and the models with two or four derivatives more are special
and cannot be obtained by any limiting procedure from the general results which hold for
higher-derivative regulated actions, which contain six or more derivatives than in the minimal renormalizable model. For the case of QG in $d=4$ in the minimal model we have obviously four derivatives. Of course, this discontinuity is related to the different type of enhanced
renormalizability properties of the models in question. As we have already explained above the gravitational model
with six derivatives in $d=4$ is the first super-renormalizable model of QG, where from the fourth loop on the
perturbative UV divergences are completely absent. The case of Stelle theory gives just the
renormalizable theory, where the divergences are present at any loop order (they are always the same
divergences, always absorbable in the same set of counterterms since the theory is renormalizable). One
sees the discontinuity already in the behaviour of UV divergences as done in the analysis of power
counting. When the number of derivatives is increased in steps (by two), then the level of loops when one
does not see divergences at all decreases but in some discontinuous jumps. And for example for the QG theory
with ten or more derivatives the UV divergences are only at the one-loop level. (For gravitational theories with
$8$ derivatives the last level which is divergent is the second loop.)

There exist also analytic formulas, which combine the results for UV divergences for the cases of theories
 with four or more derivatives more
compared to the minimal renormalizable model with four derivatives in $d=4$. Again one sees from such formulas,  that the correct
results for the minimal renormalizable model and the one with six derivatives are discontinuous. Then the case
with 8-derivative gravitational theory is the first one for which the analytic formulas hold true. However, this has apparently
nothing to do with the strengthened super-renormalizability properties at some loop level as it was emphasized above.

The six-derivative gravitational theory  is therefore 3-loop super-renormalizable since the 3-loop level is the
last one, when one needs to absorb infinities and renormalize anew the theory. These jumps from 3-loop
super-renormalizability to 2-loop and finally to one-loop super-renormalizability are from their nature
discontinuous and hence also the results for divergences inherit this discontinuity. For theories with
ten or more derivatives we have one-loop super-renormalizability and the results for even higher number
of derivatives $2n$ must be continuous in the parameter of the number of derivatives $2n$, which could be
analytically extended to the whole complex plane from the even  integer values $2n\geqslant10$, which it
originally had. In this analytically extended picture, the cases with eight, six and four derivatives are
special isolated points, which are discontinuous and cannot be obtained from the general analytic
formula valid for any $n\geqslant5$.
The origin of this is again in power counting of divergences, when some integrals over loop momenta are
said to be convergent, when the superficial degree of divergence is smaller than zero, and when this is
non-negative, then one meets non-trivial UV divergences. These infinities are logarithmic in the UV cutoff $k_{\rm UV}$ for loop integration
momenta  for the degree $\Delta$ vanishing, and power-law type for the degree $\Delta$ positive. This
sharp distinction between what is convergent and what is divergent (based on the non-negativity of the degree
of divergences $\Delta$ of any diagram) introduces the discontinuity, which is the main source of the problems
here.

In this contribution, we mainly discuss and analyze the results which were first obtained in our recent
publication \cite{Rachwal:2021bgb}. The details of the methods used to obtain them were presented to some
extent in this recent article. The method consists basically of using the Barvinsky-Vilkovisky trace
technology \cite{Barvinsky:1985an} applied to compute functional traces of differential operators giving the expression for the
UV-divergent parts of the effective action at the one-loop level. The main results were obtained in
background field method and from UV divergences in \cite{Rachwal:2021bgb} we read the beta functions of
running dimensionless gravitational couplings. The results for them in six-derivative gravitational theory in $d=4$
spacetime dimensions were the main results there. They are also described in section \ref{s2} here.
Instead, in the present contribution, we decided to include an extended discussion of the theoretical
checks done on these results in section \ref{s3}. However, the main novel contribution is in section
\ref{s4}, where we present the analysis of the structure of these obtained results for the beta functions.
Our main goal here is to show an argumentation that provides an explanation why the structure of the beta
function is unique and why it depends in this particular form on the ratio $x$ (to be defined later in
the main text in (\ref{xdefinition})). These comments were not initially included in the main research
article \cite{Rachwal:2021bgb} and they constitute the main new development of the present paper.

We remind to the reader that in this paper, in particular, we will spend some time on attempts to explain the
discontinuity of such results for UV divergences, when one goes from six- to four-derivative gravitational
theories. So, in other words, when one reduces 3-loop super-renormalizability to just renormalizability.
Or equivalently, when the situation at the fourth perturbative loop gets modified from not having divergences at all,
because all loop integrations give convergent results (with negative superficial degree $\Delta<0$), to the situation
when at this loop level still UV divergences are present (since their degree $\Delta$ is zero for logarithmic UV
divergences in the cutoff). This clearly sharp contrast in the sign of the superficial degree of
divergences is one of the reasons, why the discontinuity between the cases of six- and four-derivative
gravitational theories in $d=4$ persists.

\subsection{Addition of killer operators}
\label{s1.2}
As a matter of fact, we can also add other terms (cubic in gravitational curvatures ${\cal R}^3$) to the Lagrangian in (\ref{lagr}). These terms again will come with the coefficients of the highest energy dimensionality, equal to the dimensionality of the coefficients $\omega_C$  and $\omega_R$. Hence they could contribute to the leading four-derivative terms with UV divergences of the theory. The general form of them is given by the following list of six GR-covariant terms
\begin{multline}
{\cal L}_{{\cal R}^3}=s_1R^{3}+s_{2}RR_{\mu\nu}R^{\mu\nu}+s_{3}R_{\mu\nu}R^{\mu}{}_{\rho}R^{\nu\!\rho}\\
 +s_4RR_{\mu\nu\!\rho\sigma}R^{\mu\nu\!\rho\sigma}+s_{5}R_{\mu\nu}R_{\rho\sigma}R^{\mu\!\rho\nu\sigma}\\
 +s_6R_{\mu\nu\!\rho\sigma}R^{\mu\nu}{}_{\kappa\lambda}R^{\rho\sigma\kappa\lambda}\,.
 \label{killers}
\end{multline}
Actually, these terms can be very essential for making the gravitational theory with six-derivative actions completely UV-finite. However, for renormalizability or super-renormalizability properties these terms are not necessary, e.g., they do not make impact on the renormalizability of the theory and therefore should be regarded as non-minimal. In the analysis below we did not take their contributions into account and made already a technically demanding computation in a simplest minimal model with six-derivative actions. The set of terms in (\ref{killers}) is complete in $d=4$ for all what regards terms cubic in gravitational curvatures. This non-trivial statement is due to various identities as proven in \cite{Mistry:2020iex}.

These cubic terms are also sometimes called ``killers'' of the beta functions since  they may have profound effects on the form of the beta functions of all terms in the theory. This is roughly very simple to explain. These killer terms are generally of the type $s{\cal R}^3$ and are to be added to the original Lagrangian in (\ref{lagr}) of six-derivative theories, where the UV-leading terms were of the type $\omega {\cal R}\square{\cal R}$. It is well known that to extract UV divergences at the one-loop level one has to compute the second variational derivative operator (Hessian $\hat H$) from the full action. The contributions from cubic killers to it will be of the form of at least $s{\cal R}$, when counted in powers of generalized curvature ${\cal R}$. Next, when computing the trace of the functional logarithm of the Hessian operator for the form of the one-loop UV-divergent effective action one uses the expansion of the logarithm in a series according to
\begin{equation}
\ln(1+z)=z-\frac12 z^2+\ldots.
\label{lnexp}
\end{equation}
Hence we need to take maybe up to the square of the contribution $s{\cal R}$ to the Hessian from the cubic killer term. The third power would be too much. We must remember that we are looking for terms of the general type ${\cal R}^2$  in the UV-divergent part of the effective action. Hence the contribution of the cubic killer in curvatures would produce addition to the covariant terms with UV divergences of the general type $f(s){\cal R}^2$, where the yet unknown functions $f(s)$ can be polynomials up to the second order in the coefficients $s_i$ of these killers. Now, requiring the total beta functions vanish (for complete UV-finiteness)  we need in general to solve the system of the quadratic equations in the coefficients $s_i$. The only obstacle for finding coefficients of the killers can be that some solutions of this system reveal to be complex numbers, not real, but we need to require all $s_i$ coefficients to be real for the definiteness of the action (for example in the Euclidean case of the signature of the metric). Therefore this issue requires a more detailed mathematical analysis, but the preliminary results based on \cite{Modesto:2014lga,Rachwal:2021bgb} show that in most of the cases the UV-finiteness is possible and easily can be achieved by adding the cubic killer operators from (\ref{killers}) with real coefficients $s_i$.

One can compare the situation here with cubic killers to the more known situation where the quartic killers are used to obtain UV-finiteness. Unfortunately, such quartic killers cannot be added to the six-derivative gravitational theory from (\ref{lagr}) since they would have too many partial derivatives and would destroy the renormalizability of the model. Quartic killers can be included in theories with at least 8 derivatives. Such approach seems to be preferred one since the contribution of quartic killers (of the type schematically as ${\cal R}^4$) is always linear in $d=4$ to UV divergences proportional to ${\cal R}^2$ schematically. And to solve linear system of equations with linear coefficients is always doable and one always finds solutions and they are always real. This approach was successfully applied to gravity theories in \cite{Modesto:2014lga}, to gauge theories in \cite{Modesto:2015foa}, to the theories on de Sitter and anti-de Sitter backgrounds \cite{Koshelev:2017ebj} and also in general non-local theories \cite{Modesto:2017sdr}. One could show that the UV-finiteness may be an universal feature of quantum field-theoretical interactions in nature \cite{Modesto:2015lna}. Moreover, this feature of the absence of perturbative UV divergences is related to the quantum conformality as advocated in \cite{Modesto:2016max,Modesto:2017xha}.

\subsection{Universality of the results}
\label{s1.3}

Finally, one of the most important features of the expression for the UV-divergent part of the effective action in the six-derivative gravitational theories is its complete independence of any parameter used in the computation. This parameter can be gauge-fixing parameter, or it can appear in gauge choice, or in details of some renormalization scheme, etc. This bold fact of complete universality of the results for the effective action was proven by the theorem by Kallosh, Tarasov and Tyutin (KTT) \cite{Kallosh:1978wt,Kalmykov:1995fd,Shapiro:1994ww}, applied here to the six-derivative QG theories. The theorem expresses the difference between two effective actions of the same theory but computed using different set of external parameters. Basically, this difference is proportional to the off-shell tensor of classical equation of motion of the original theory. And this difference disappears on-shell. However, in our computation we want to exploit the case when the effective action and various Green functions are computed from it understood as the off-shell functional.

But in super-renormalizable theories there is still some advantage of using this theorem, namely for this one notices the difference in number of derivatives  on the metric tensor between the original action and Lagrangian of the theory as it is in the form (\ref{sred}) (and resulting from it classical EOM) and between the same counting of derivatives done in the divergent part of the effective action. We remind the reader that in the former case we have six derivatives on the metric, while in the latter we count up to four derivatives. This mismatch together with the theorem of  KTT implies that the difference between the two UV-divergent parts of the effective actions (only for these parts of the effective actions) computed using two different schemes or methods must vanish in super-renormalizable QG theories with six-derivative actions for whatever change of the external parameters that are used for the computation of these UV-divergent functionals. This means that our results for divergences are completely universal and cannot depend on any parameter. Hence we derive the conclusion that our found divergences do not depend on the gauge-fixing parameters, gauge choices nor on other parametrization ambiguities. We remark that this situation is much better than for example in E-H gravity, where the dependence on a gauge is quite strong, or even in Stelle four-derivative theory, where four-derivative UV-divergent terms also show up some ambiguous dependence on  gauge parameters off-shell. Here we are completely safe from such problems and such cumbersome ambiguities.

In this way such beta functions are piece of genuine observable quantity that can be defined in super-renormalizable models of QG. They are universal, independent of spurious parameters needed to define the gauge theory with local symmetries, and moreover they are exact, but still being computed at the one-loop level in perturbation calculus. They are clearly  very good candidates for the observable in QG models. Therefore all these nice features gives us even more push towards analyzing the structure of such physical quantities and to understand this based on some theoretical considerations. This is what we are trying to attempt in this contribution.

Another important feature is that in theories with higher derivatives in their defining classical action,
on the full quantum level there is no need for perturbative renormalization of the graviton`s wave
function. This is also contrary to the case of two-derivative theory, when one has to take this phenomena
into account, although its expression is not gauge-invariant and depends on the gauge fixing. These nice
properties of no need for wave function renormalization can be easily understood in the
Batalin-Vilkovisky formalism for quantization of gauge theories (or in general theories with differential
constraints) \cite{Batalin:1983ggl,Batalin:1981jr}. This important feature is also shared by other, for example, four-derivative QG
models. Since the wave function of the graviton does not receive any quantum correction, then one can
derive the form of the beta functions for couplings just from reading the UV divergences of the dressed
two-point functions with two external graviton lines. We can simplify our computation drastically since
for this kind of one-loop computation we do not have to bother ourselves with the three- or higher
$n$-point function to independently determine the wave function renormalization. Unfortunately, the latter
is the case, for example, for standard gauge theory (Yang-Mills model) or for E-H gravity, where the
renormalization of the coupling constant of interactions has to be read from the combination of the two-
and three-point functions of the quantum theory, while the wave function renormalization of gauge fields
or graviton field respectively can be just read from quantum dressed two-point Green function. For the
case of six-derivative theories, just from the two-point function we can read everything about the
renormalization of the coupling parameters of gravitons' interactions. Additionally, we have that on the first quantum loop level
we do not need to study effective interaction vertices dressed by quantum corrections. Hence, here at the
one-loop level there is no quantum renormalization of the graviton's wave function and UV divergences
related to interactions are derived solely from propagation of free modes (here of graviton fields) around
the flat spacetime and corrected (dressed) at the first quantum loop. Effective vertices of interactions
between gravitons do not matter for this, but that situation may be changed at higher loop orders. At the
one-loop level this is a great simplification for our algorithm of derivation of the covariant form of UV
divergences since we just need to extract them from the expression for one-loop perturbative two-point
correlators of the theory, both in cases of four- and six-derivative QG models.

All these nice features of the six-derivative QG model makes it further worth studying as an example of
non-trivial RG flows in QG. Here we have exactness of one-loop expressions for running $\theta_C(t)$,
$\theta_R(t)$ and $\theta_{\rm GB}(t)$ coupling parameters in (\ref{lagr}), together with
super-renormalizability. This is one of the most powerful and beautiful features of the
super-renormalizable QG theory analyzed here. Therefore, this model gives us a good and promising theoretical
laboratory for studying RG flows in general quantum gravitational theories understood in the
field-theoretical framework.

We remark that from a technical point of view, the one-loop calculations in super-renormalizable models of QG are more difficult when
compared to the ones done in the four-derivative just renormalizable gravitational models \cite{Fradkin:1981iu,Barvinsky:1985an,Avramidi:1985ki,Brown:1977pq}. The level of complexity of such  calculations
depends strongly on the number of derivatives in the classical action of the model as well as on the type of one-loop counterterms one is looking for. The counterterm for the cosmological constant is actually very easy to obtain and this was done already in \cite{Asorey:1996hz}. Next, the derivation of the divergence linear in the scalar curvature $R$ requires really big efforts and was achieved only recently in our collaboration in \cite{Modesto:2017hzl}. In the present work, we comment on  the next step, and we show the results of the  calculations of the simply looking one-loop UV divergences for
the four-derivative sector in the six-derivative minimal gravity model. In our result, we have now full answers to the beta functions for the Weyl-squared $C^2$, Ricci scalar-squared $R^2$ and the Gauss-Bonnet ${\rm GB}$ scalar terms. The calculation is really tedious and cumbersome and it was done for the simplest possible six-derivative QG theory without
cubic terms in the classical action, which here would be third powers of the generalized curvature tensor ${\cal R}^3$. Even in
this simplest minimal case, the intermediate expressions are too large for the explicit presentation here, hence they will be mostly omitted.
Similar computations in four-, six- and general higher-derivative gauge theory were also performed in 
\cite{Asorey:1995tq,Modesto:2015foa,Asorey:2020omv}.

As it was already mentioned above, the derivation of zero- and two-derivative ultraviolet divergences
has been previously done in Refs. \cite{Asorey:1996hz} and \cite{Modesto:2017hzl}. Below  we will show the results for the complete set of beta functions for the
theory (\ref{sred}). This we will achieve  by deriving the exact and computed at one-loop beta function coefficients for the four-derivative gravitational couplings, namely $\theta_C$, $\theta_R$ and $\theta_{\rm GB}$, extracted as the coefficients of the UV-divergent part of the effective action in (\ref{sdiv}). Without loss of
generality, the calculation will be performed in the reduced model  (\ref{sred}), so without terms subleading in the number of partial derivatives acting on the metric tensor after the proper expansion here. (We will not need to include terms like $R^2$, $C^2$  or even $R$ in (\ref{sred}).) This is clearly explained by the arguments from dimensional analysis since the divergences with four derivatives of the metric, in (\ref{sdiv}), are of our biggest interest here. Moreover, numerical coefficients of those subleading terms cannot in any way combine with coefficients of propagators (shaped in the UV regime by the leading terms with six derivatives in the action (\ref{sred})) to form dimensionless ratios in front of terms in (\ref{sdiv}) in $d=4$ spacetime dimensions.

\section{Brief description of the technique for computing UV divergences}
\label{s2}

An essential part of the calculations is pretty much the same as
usually done in any higher-derivative QG model, especially in the renormalizable or super-renormalizable models
\cite{Modesto:2017hzl,Rachwal:2021bgb} as considered here. In what follows, we can skip a great part of the explanations.
We will focus on the calculation of
the fourth derivative terms of the divergent part of the effective action.

First, to perform pure computation we use the background field method, which is defined by the following splitting of the metric
\begin{equation}
g_{\mu\nu}  \longrightarrow  {\bar{g}}_{\mu\nu} + h_{\mu\nu}
\label{backgr}
\end{equation}
to the background $\bar{g}_{\mu\nu}$ and the quantum fluctuation parts given by the spin-2 symmetric tensor $h_{\mu\nu}$.

The next step is to define the gauge-fixing condition. Since our theory with six derivatives still possesses gauge invariance due to diffeomorphism symmetry we have to fix the gauge to make the graviton propagator non-degenerate.
For this we will make some choice of the gauge-fixing parameters, here represented by numerical $\alpha$, $\beta$ and $\gamma$ parameters.
First, we choose
the parameter $\beta$ in the harmonic background gauge-fixing condition $\chi_\mu$, according to
\begin{equation}
\chi_{\mu} =
\nabla^\lambda h_{\lambda\mu} - \beta \, \nabla_\mu h, \qquad
h =  h^{\nu}{}_{\nu} \, ,
\label{chi}
\end{equation}
in the most simple ``minimal'' form, as will be indicated below. The same
concerns the parameters $\alpha$ and $\gamma$. Finally, we select a  general form of the
weighting operator, $\,\hat{C}=\tilde{C}^{\mu\nu}$, which is defined
by the formula below:
\begin{multline}
{\hat  C}=
\tilde{C}^{\mu\nu}
\,=\,
\,-\frac{1}{\alpha}
\left(g^{\mu\nu}\Box^2 + (\gamma-1)\nabla^\mu\Box \nabla^\nu\right).
\label{weight}
\end{multline}
This together with the gauge-fixing condition, that is $\chi^\mu$,
defines the gauge-fixing action \cite{Fradkin:1981iu} in the following form,
\begin{equation}
S_{{\rm gf}}=\!\int\! d^{4}x\sqrt{|g|}\,\,\chi_\mu \,\tilde{C}^{\mu\nu}\,
\chi_\nu .
\label{gf}
\end{equation}
The action of the complex Faddeev-Popov (FP) ghost fields (respectively $\bar{C}^\mu $ and $C_\mu$)  has in turn the form
\begin{equation}
S_{{\rm gh}}
\,=\,
\!\int\! d^{4}x\sqrt{|g|}\,\, \bar{C}^{\mu}M_{\mu}{}^{\nu}C_{\nu},
\label{gh}
\end{equation}
where the bilinear part between the anti-ghost $\bar{C}^\mu$ and ghost fields $C_\mu$, the so called FP-matrix $\hat M$, depends differentially on $\chi_\mu$ gauge-fixing conditions and also on the contracted form of the generator of gauge transformations $\hat R$,
\begin{equation}
{\hat  M}\,=\,
M_{\mu}{}^{\nu}
\,=\,\frac{\delta \chi_\mu}{\delta g_{\alpha\beta}}\,R_{\alpha\beta}\,^\nu
\,=\,\delta_\nu^\mu \Box
+ \nabla^\nu \nabla_\mu - 2\beta\nabla_\mu \nabla^\nu.
\label{M}
\end{equation}
In the above equation by the matrix-valued operator $R_{\alpha\beta}\,^\nu$ we mean the generator of infinitesimal diffeomorphism (local gauge) transformations in any metric theory of gravity.

Since as proven and explained at the end of section \ref{s1.3}, our final results for UV divergences are here completely universal and they are independent of any parameter used to regularize, compute and renormalize the effective action of the theory, then we can take the following philosophy at work here. We choose some specific gauge choice in order to simplify our calculation, but then we are sure that the final results will be still correct, if obtained consistently within this computation done in a particular gauge choice. It is true that intermediate steps of the computation may be different in different gauges, but the final results must be unique and it does not matter which way we arrive to them. We think we could choose one of the simplest path to reach this goal. A posteriori this method is justified, but the middle steps of the processing of the Hessian operator will not have any invariant objective physical meaning. These are just steps in the calculational procedure in some selected gauge.

One knows that in such a case, for example, for a formalism due to Barvinsky-Vilkovisky (BV) \cite{Barvinsky:1985an} of functional traces of differential operators applied in the background field method framework, all intermediate results are manifestly gauge-independent. Then still such partial contributions (any of them) separately do not have any sensible physical meaning, although such results are gauge-independent and look superficially physical -- any physical meaning cannot be properly associated to them, if all these terms are not taken in total and only in the final sum. On the contrary, if the computation is performed using Feynman diagrams, momentum integrals and around flat spacetime, then the intermediate results are not gauge-invariant, as it is well known for partial contributions of some graphs, and only in the final sum they acquire such features of gauge-independence.

We also need to distinguish here two different features. Some partial results may be still gauge-dependent and their form may not show up gauge symmetry (for example, using Feynman diagram approach, a contribution from a subset of divergent diagrams may not be absorbed by a gauge-covariant counterterm: $F^2$ in gauge theories, or ${\cal R}^2$ in gravity in $d=4$). This feature should be however regained when the final results are obtained. This is actually a good check of the computation. But another property is independence of the gauge-fixing parameters, which are spurious non-physical parameters. At the same time, a counterterm might be gauge-covariant (built with $F^2$  or ${\cal R}^2$ terms), but its front coefficient may depend on these gauge parameters $\alpha$, $\beta$, $\gamma$, etc. This should not happen for the final results and they should be both gauge-covariant (so gauge-independent or gauge-invariant) and also gauge-fixing parameters independent. These two necessary properties, to call the result physical, must be realized completely independently and they are a good check of the correctness of the calculation.

Unfortunately, it seems that using the BV computational methods even in the intermediate results for traces of separate matrix-valued differential operators (like $\hat H$, $\hat M$ and $\hat C$), we see already both gauge-independence and gauge-fixing parameters independence provided that such parameters were not used in the definition of these operators. Only in some cases, the total result is only gauge-fixing parameter independent. This means that within this formalism of computation this check is not very valuable and one basically has to be very careful to get the correct results at the end. Instead, we perform a bunch of other rigorous checks of our results as it is mentioned, for example, in section \ref{s3}.

Finally, let us here give briefly a few details concerning the choice of the
gauge-fixing parameters $\alpha$, $\beta$ and $\gamma$.  The bilinear form of the action is defined from the
second variational derivative (giving rise to the Hessian operator $\hat H$)
\begin{equation}
{\hat H}
=
H^{\mu\nu,\rho\sigma}
=
\frac{1}{\sqrt{|g|}}\,\,
\frac{\delta^2 \left(S 
+ S_{{\rm gf}}
\right)}{\delta h_{\mu\nu}\,\delta h_{\rho\sigma}}
\,=\,H_{{\rm lead}}^{\mu\nu,\rho\sigma} \,+\,
{\cal O}(\nabla^4),
\label{bil}
\end{equation}
where the first term $H_{{\rm lead}}^{\mu\nu,\rho\sigma}$
contains six-derivative terms, which are leading in the UV regime. By
$\,{\cal O}(\nabla^4)\,$ we denote  the rest of the bilinear form, with four or
less derivatives and with higher powers of gravitational curvatures $\cal R$.
 The energy dimension of this expression is compensated by the powers of
curvature tensor $\cal R$ and its covariant derivatives, hence in this case, we can also
denote $\,{\cal O}(\nabla^4)\,=\,O({\cal R})$. The
corresponding full expression for the Hessian operator $\hat H$ is very bulky, and we will not include it here.

The highest derivative part (leading in the UV regime) of the ${\hat H}$ operator, after
adding the gauge-fixing term (\ref{gf}) that we have selected, has the form
\begin{multline}
H_{{\rm lead}}^{\mu\nu,\rho\sigma}=
\Big[ \omega_C\,\delta^{\mu\nu , \rho\sigma}
\,+\, \Big( \frac{\beta^{2}\gamma}{\alpha}
- \frac{\omega_C}{3}
+ 2\omega_{R}\Big) g^{\mu\nu}g^{\rho\sigma}\Big] \square^3
\\
+
\Big( \frac{\omega_C}{3} - 2\omega_R - \frac{\beta\gamma}{\alpha}\Big)
\Big( g^{\rho\sigma}\nabla^{\mu}\nabla^{\nu}
+ g^{\mu\nu}\nabla^{\rho}\nabla^{\sigma} \Big) \square^{2}
\\
+
\Big( \frac{1}{\alpha}g^{\mu\rho} - 2{\omega_C} g^{\mu\rho} \Big)
\nabla^{\nu}\nabla^{\sigma}\square^{2}
\\
+ \Big( \frac{2\omega_C}{3} + 2\omega_{R} + \frac{\gamma-1}{\alpha}\Big)
\nabla^{\mu}\nabla^{\nu}\nabla^{\rho}\nabla^{\sigma}\square .
\label{bilin}
\end{multline}
In this expression, we do not mark explicitly the symmetrization in and between the pairs of indices
$(\mu,\nu)$ and $(\rho,\sigma)$ for the sake of brevity.

To make the UV-leading part of the Hessian operator $H_{{\rm lead}}^{\mu\nu,\rho\sigma}$ minimal, one
has to choose the following values for the gauge-fixing parameters
\cite{Modesto:2017hzl}:
\begin{equation}
\alpha=\frac{1}{2\omega_{C}},
\qquad
\beta=\frac{\omega_{C}-6\omega_{R}}{4\omega_{C}-6\omega_{R}},
\qquad
\gamma=\frac{2\omega_{C}-3\omega_{R}}{3\omega_{C}}.
\label{mingauge}
\end{equation}

We previously explained that this choice does not affect the values and the form of one-loop divergences in super-renormalizable
QG. Thus, we assume it as the most simple option.

One notices that the expressions for gauge-fixing parameters in (\ref{mingauge}) are singular in the limit $\omega_C\to0$ and also when $\omega_C=\frac{3}{2}\omega_R$. While the first one is clearly understandable, because then we are losing one term $\omega_CC\square C$ in the action (\ref{sred}) and the theory is degenerate and non-generic, the second condition is not easily understandable in the Weyl basis of writing terms in the action in (\ref{sred}) (with $R^2$  and $C^2$ terms). To explain this other spurious degeneracy one rather goes to the Ricci basis of writing terms (with $R^2$  and $R^2_{\mu\nu} = R_{\mu\nu}R^{\mu\nu}$ elements and also properly generalized to the six-derivative models by inserting one power of the box operator in the middle). There one sees that the absence of the coefficient in front of the $R^2_{\mu\nu}$ leads to the pathology in the case of $\omega_C=\frac{3}{2}\omega_R$ and also formal divergence of the $\beta$ gauge-fixing parameter. We remark that in the final results there is no any trace of this denominator and this divergence, hence the condition for non-vanishing of the coefficient in front of the covariant term  $R^2_{\mu\nu}$ in the Ricci basis does not have any sensible and crucial  meaning -- this is only a spurious intermediate dependence on $(4\omega_{C}-6\omega_{R})^{-1}$. Contrary, the singular dependence on $\omega_C$  coefficient is very crucial and will be analyzed at length here. Actually, to verify that the denominators with $(4\omega_{C}-6\omega_{R})^{-1}$ completely cancel out in the final results is a powerful check of our method of computation.

Now we can collect all the necessary elements to write down the general
formula for the UV-divergent part of the one-loop contribution to the effective action of
the theory \cite{Fradkin:1981iu},
\begin{equation}
{\bar \Gamma}^{(1)}
\,=\,\frac{i}{2} {\rm Tr} \ln {\hat H}
\,-\,i {\rm Tr}\ln  {\hat  M}
\,-\, \frac{i}{2}{\rm Tr} \ln  {\hat C}.
\label{TrLn}
\end{equation}

The calculation of the divergent parts of the first two expressions
in (\ref{TrLn}) is very standard. One uses for this the technique of the generalized
Schwinger-DeWitt method \cite{Barvinsky:1985an}, which was first introduced by Barvinsky and Vilkovisky.
For this reason we
shall skip most of the standard technical details here.
We use the Barvinsky-Vilkovisky trace technology related to the covariant heat kernel methods together with methods of dimensional regularization (DIMREG) to evaluate the functional traces present in (\ref{TrLn}) and to have under control the general covariance of the final results. Due to this we cannot check it because all three contributions in (\ref{TrLn}) gives results which look covariant and sensible. We remind the reader that here we work with the minimal gauge choice and in general all three terms separately will show the gauge dependence and also spurious dependence on gauge-fixing parameters $\alpha$, $\beta$ and $\gamma$. However, only the final results, so the weighted sum as in (\ref{TrLn}) is properly gauge-independent and gauge-fixing independent and gives rise to a physical observable of the beta functional of the theory at the one-loop level.

The computational method that we adopt here consists basically of using the Barvinsky-Vilkovisky trace technology to compute functional traces of differential operators giving the expression for the UV-divergent parts of the effective action at the one-loop level. The main results are obtained in background field method and from UV divergences in \cite{Rachwal:2021bgb} we read the beta functions of running gravitational couplings. We also present here below an illustrative scalar example of the techniques by which these results were obtained.

\subsection{Example of the BV method of computation for the scalar case}
\label {s2.1}

The simplest example to use the technique of computation presented here can be based on the analysis of the scalar case given by the action
\begin{equation}
S=\!\int\! d^{4}x\left(-\frac{1}{2}\phi\square\phi-\frac{\lambda}{4!}\phi^{4}\right).
\label{scalaraction}
\end{equation}
From this action one reads the second variational derivative operator (also known as the Hessian) given by the formula
\begin{equation}
H=\frac{\delta^{2}S}{\delta\phi^{2}}=-\square-\frac{\lambda}{2}\phi^{2}.
\end{equation}
Next, one needs to compute the following functional trace ${\rm Tr}\ln H$ to get the UV-divergent part of the one-loop effective action
\begin{multline}
{\rm Tr}\ln H={\rm Tr}\ln\left(-\square-\frac{\lambda}{2}\phi^{2}\right)\\
={\rm Tr}\ln\left(-\square\left(1+\frac{\lambda}{2}\phi^{2}\square^{-1}\right)\right)\\
={\rm Tr}\ln\left(-\square\right)+{\rm Tr}\ln\left(1+\frac{\lambda}{2}\phi^{2}\square^{-1}\right).
\end{multline}
In the above expression, one concentrates on the second part which contains the $\lambda$ coupling. One expands the logarithm, as in (\ref{lnexp}),  in the second trace to the second order in $\lambda$. This yields
\begin{multline}
{\rm Tr}\ln\left(1+\frac{\lambda}{2}\phi^{2}\square^{-1}\right)={\rm Tr}\left(\frac{\lambda}{2}\phi^{2}\square^{-1}\right)\\
-\frac{1}{2}{\rm Tr}\left(\frac{\lambda}{2}\phi^{2}\square^{-1}\right)^{2}+\ldots
\end{multline}
and one picks up from it only the expression quadratic in $\lambda$ and quartic in the background scalar field $\phi$, which is also formally quadratic in the inverse box operator $\square^{-1}$, that is the part
\begin{equation}
{\rm Tr}\ln H\supset-\frac{1}{2}\frac{\lambda^{2}}{4}\phi^{4}{\rm Tr}\,\square^{-2}=-\frac{\lambda^{2}}{8}\phi^{4}{\rm Tr}\,\square^{-2}.
\end{equation}
Precisely this expression is relevant for the UV divergence proportional to the quartic interaction term $-\frac{\lambda}{4!}\phi^{4}$ in the original scalar field action (\ref{scalaraction}). Noticing that the functional trace of the $\square^{-2}$ scalar operator in $d=4$ is given by
\begin{equation}
{\rm Tr}\,\square^{-2}=i\frac{\ln L^{2}}{(4\pi)^{2}},
\end{equation}
where $L$ is a dimensionless UV-cutoff parameter related to the $\Lambda_{{\rm UV}}$ dimensionful momentum UV-cutoff and the renormalization scale $\mu$ via $\Lambda_{{\rm UV}}=L\mu$, one finds for the UV-divergent and interesting us part of the one-loop effective action here
\begin{equation}
\Gamma_{{\rm div}}^{(1)}=\frac{i}{2}{\rm Tr}\ln H\supset\!\int\! d^{4}x\frac{\ln L^{2}}{(4\pi)^{2}}\frac{\lambda^{2}}{16}\phi^{4}.
\end{equation}

Now, one can compare this to the original action terms in (\ref{scalaraction}) describing quartic interactions of the scalar fields $\phi$: $-\!\int\! d^{4}x\frac{\lambda}{24}\phi^{4}$. The counterterm action (to absorb UV divergences) is opposite to $\Gamma_{{\rm div}}$ and the form of the terms in the counterterm action is expressed via perturbative beta functions of the theory. That is in the counterterm action $\Gamma_{{\rm ct}}$ we expect terms
\begin{equation}
\Gamma_{{\rm ct}}=-\Gamma_{{\rm div}}=-\frac{1}{2}\frac{\ln L^2}{24}\!\int\! d^{4}x\beta_{\lambda}\phi^{4}
\end{equation}
with the front coefficient exactly identical to the one half of the one in front of the quartic interactions in the original action in (\ref{scalaraction}) (being equal to $-\frac{1}{4!}=-\frac{1}{24}$). From this one reads that (identifying that effectively $\ln L^2\to1$ for comparison)
\begin{equation}
-\frac{1}{48}\beta_{\lambda}=-\frac{\lambda^{2}}{16(4\pi)^{2}}
\end{equation}
and finally that
\begin{equation}
\beta_{\lambda}=\frac{3\lambda^{2}}{(4\pi)^{2}},
\end{equation}
which is a standard result for the one-loop beta function of the quartic coupling $\lambda$ in $\frac{1}{4!}\lambda\phi^{4}$ scalar theory in $d=4$ spacetime dimensions.

One sees that even in the simplest framework, the details of such a computation are quite cumbersome, and we decide not to include in this manuscript other more sophisticated illustrative examples of such derivation of the explicit results for beta functions of the theory. The reader, who wants to see some samples can consult more explicit similar calculations as presented in references \cite{Rachwal:2021bgb,Modesto:2017hzl,Modesto:2015foa}. In particular, the appendix of \cite{Modesto:2015foa} compares two approaches to the computation of UV divergences in gauge theory (simpler than gravity but with non-Abelian gauge symmetry) -- using BV heat kernel technique and using standard Feynman diagram computation using graphs and Feynman rules around flat space and in Fourier momentum space.

\subsection{Results in six-derivative gravity}
\label{s2.2}
The final results for this computation of all UV divergences of the six-derivative gravitational theory are

\begin{multline}
\Gamma^{(1) R, C}_{\rm div}
=-\frac{\ln L^{2}}{2(4\pi)^{2}}\!\int\! d^4x \sqrt{|g|}
\left\{
\Big(\frac{397}{40}+\frac{2x}{9}\Big)C^2\right.\\\left.
+ \frac{1387}{180}{\rm GB}
- \frac{7}{36}R^{2}
 \right\}
\qquad
\label{sixdergamma}
\end{multline}
for the case of six-derivative pure QG model in $d=4$ spacetime dimensions
and

\begin{multline}
\Gamma^{(1) R, C}_{\rm div}
=
-\frac{\ln L^{2}}{2(4\pi)^{2}}\!\int\! d^4x \sqrt{|g|}
\left\{
-\frac{133}{20}C^2\right.\\\left.
+ \frac{196}{45}{\rm GB}
+\left( -\frac52 x_{\rm4-der}^{-2}+\frac52 x_{\rm4-der}^{-1}-\frac{5}{36}\right) R^{2}
 \right\}
\qquad
\label{fourdergamma}
\end{multline}
for the case of four-derivative pure Stelle quadratic model of QG to the one-loop accuracy. This last result was first reported in \cite{Avramidi:1985ki}. The result in six-derivative gravity is freshly new \cite{Rachwal:2021bgb}. Here we define the covariant cut-off regulator $L$
\cite{Barvinsky:1985an}, which stays in the following relations to the dimensional regularization parameter $\epsilon$
\cite{Brown:1977pq,Barvinsky:1985an},
\begin{equation}
\ln L^2 \equiv \ln \frac{ \Lambda_{\rm UV}^2 }{\mu^2}
\,=\, \frac{1}{\epsilon} = \frac{1}{2 - \omega}
=  \frac{2}{4-n}\,,
\label{dimreg}
\end{equation}
where we denoted by $n$ the generalized dimensionality of spacetime in the DIMREG scheme of regularization (additionally $\Lambda_{\rm UV}$ is the dimensionful UV cutoff energy parameter and $\mu$ is the quantum renormalization scale).
Moreover, to write compactly our finite results for the six-derivative theory we used the definition of the fundamental ratio of the theory $x$ as
\begin{equation}
x=\frac{\omega_C}{\omega_R},
\label{xdefinition}
\end{equation}
while for Stelle four-derivative theory in (\ref{fourdergamma}) we use analogously but now with the theta couplings instead of omegas, namely
\begin{equation}
x_{\rm4-der}=\frac{\theta_C}{\theta_R}.
\end{equation}

It is worth to describe briefly here also the passage from UV divergences of the theory at the one-loop level to the perturbative one-loop beta functions of relevant dimensionless couplings. Using the divergent contribution to the quantum effective action,
derived previously, we can define the beta functions
of the theory. Let us first fix some definitions.

The renormalized Lagrangian $\mathcal{L}_{\rm ren}$ is obtained
starting from the classical Lagrangian written in terms of the
renormalized coupling constants and then adding the counterterms
to subtract the divergences,
\begin{multline}
 \mathcal{L}_{\rm ren} =  \mathcal{L}(\alpha_{b}(t))
 \,=\,  \mathcal{L}\left(  Z_{\alpha_i(t)}  \alpha_i(t) \right)
\,=\, \mathcal{L}(\alpha_i(t)) + \mathcal{L}_{\rm ct}
\\
 =  \mathcal{L}(\alpha_i(t))
+
\left( Z_C -1 \right) \theta_C (t) \, C^2
  +\left( Z_R -1 \right) \theta_R(t) \, R^2\\
  + \left( Z_{\rm GB} -1 \right) \theta_{\rm GB}(t)\,{\rm GB},
  \label{counterterms}
\end{multline}
where we have that
$\mathcal{L}_{\rm ct} = - \mathcal{L}_{\rm div}$ and
$\alpha_i(t) = \{ \theta_C (t),
\theta_R(t), \theta_{\rm GB}(t) \}.$ Above we denoted by $\alpha_b(t)$ the RG running bare values of coupling parameters, by ${\cal L}_{\rm ct}$ and ${\cal L}_{\rm div}$ the counterterm and divergent Lagrangians respectively, by $Z_{\alpha_i}(t)$ renormalization constants for all dimensionless couplings and finally by $\alpha_i(t)$ these running couplings. Here and above we neglect writing terms which are UV-divergent but subleading in the number of derivatives in the UV regime.
From (\ref{counterterms}), the full counterterm
action reads, already in dimensional regularization,
\begin{multline}
\Gamma^{(1)}_{\rm ct}
= - \Gamma^{(1)}_{\rm div}
= \frac{1}{2 \epsilon} \frac{1}{(4\pi)^{2}}\!\int\! d^4x \sqrt{|g|}
\bigg\{
\Big(\frac{397}{40}+\frac{2x}{9}\Big)C^2 \\
- \frac{7}{36}R^{2}
+ \frac{1387}{180}{\rm GB}\bigg\}\\
\equiv
\frac{1}{2 \epsilon} \frac{1}{(4\pi)^{2}}\!\int\! d^4x \sqrt{|g|} \left\{
\beta_C C^2 + \beta_R R^2 + \beta_{\rm GB} {\rm GB}
\right\} \, .
\end{multline}
Comparing the last two formulas we can identify the beta functions and finally get the renormalization group equations for the six derivative theory,
\begin{equation}
\beta_C = \mu\frac{d \theta_C}{d \mu}  
 = \frac{1}{(4\pi)^{2}} \left( \frac{397}{40}+\frac{2x}{9} \right),
\end{equation}
\begin{equation}
\beta_{R} = \mu\frac{d \theta_{R}}{d \mu}
= -  \frac{1}{(4\pi)^{2}} \,\frac{7}{36} \, ,
\end{equation}
\begin{equation}
\beta_{\rm GB} = \mu\frac{d \theta_{\rm GB}}{d \mu}
= \frac{1}{(4\pi)^{2}} \,\frac{1387}{180} \, ,
\label{betaGB}
\end{equation}

The three lines above constitute the main results of this work. Their structure, mainly the $x$-dependence is the main topic of discussion in the next sections. Above we denoted by $t$ the so called logarithmic RG time parameter related in the following way: $t=\log\frac{\mu}{\mu_0}$ to the renormalization scale $\mu$, where $\mu_0$  is some reference energy scale.

As we will show below the differences between the cases of four-derivative theory and six-derivative one are significant and the dependence on the ratio $x$ is with quite opposite pattern and in completely different sectors of ultraviolet divergences of the two respective theories. In the main part of this contribution we will try an attempt to explain the
mentioned difference, which is now clearly noticeable, using some general principles and arguments about renormalizability of the quantum models. We will also study some limiting cases of the non-finite (infinite or zero) values of the $x$ parameter and motivate that in such cases the QG model is non-renormalizable and this leads to characteristic patterns in the structure of beta functions mentioned above for six-derivative theories. This is also why we can call the ratio $x$ as the fundamental parameter of the gravitational theory.


\section{Some theoretical checks of the results (\ref{sixdergamma})}
\label{s3}

Let us say that regardless of the simplicity of the final formulas with the final result in \cite{Rachwal:2021bgb}, the intermediate
calculations were quite big and this is why we cannot present these intermediate steps here.  This was not only because of the size of the algebraic expressions, where we used Mathematica for help with symbolic algebra manipulations, but also
due to the complexity of all the steps of the computation starting from the quadratic expansions of the action of six-derivative classical theory. The ultimate validity of the calculations has
been checked in several different ways. This is also briefly described below.

The following checks were performed to ensure the correctness of the intermediate results of  the computation of UV divergences, which was the main task of the work presented here.
\begin{enumerate}
\item First, the validity of the expression for the Hessian operator  from  the classical  action with six derivatives was verified in the following way. The covariant divergence of the second variational derivative operator (Hessian) with respect to gravitational fluctuations $h_{\mu\nu}$, from
each GR-covariant term $S_{{\rm grav},i}$ in the gravitational action must be separately zero, namely
\begin{equation}
\nabla_{\mu}\left(\frac{\delta^{2}S_{{\rm grav},i}}{\delta h_{\mu\nu}\delta h_{\rho\sigma}}\right)=0+O\left(\nabla^{k}{\cal R}^{l},k+2l>4\right), \end{equation}
where $S_{{\rm HD}}=\sum_i S_{{\rm grav},i}$. This formula was explicitly checked for each term in the action in (\ref{sred}) to the order quadratic in curvatures and up to total of four covariant derivatives acting on the general gravitational curvature $\cal R$.

\item The computation of the functional trace of the logarithm of the gauge weighting operator $\hat C$, namely of ${\rm Tr}\ln \hat{C}$ was checked using three methods. Since the
$\hat C$ operator is a non-minimal four-derivative differential operator and matrix-valued (so with vector indices),
then the computation of its trace of the logarithm is a bit troublesome. One has to be more careful here. Therefore, we performed additional verifications of our partial results for this trace. Our three methods consist
basically of transforming the problem to computing the same trace of logarithm but of new operators (with higher number of derivatives). Next, by selecting some
adjustable parameters present in the construction of these new operators, these morphed operators could be put into a minimal form and easily traced (under the functional logarithm
operation) using standard methods and prescriptions of Barvinsky-Vilkovisky trace technology \cite{Barvinsky:1985an}. This construction of new operators was achieved by an operatorial multiplication by some two-derivative spin-one
 operator $\hat Y$ containing one free adjustable parameter.  For details one can look up the section III of \cite{Rachwal:2021bgb}.

 In the first variant of the method, we multiplied $\hat C$ from the right by $\hat Y$ one time, in the second method we multiplied by $\hat Y$
from the left also once, and in the final third method we used the explicitly symmetric form of multiplication
$\hat Y\!\hat{C}\hat Y$. (This last form of multiplication is presumably very important for the manifest self-adjointness property of the resulting 8-derivative differential operator $\hat Y\!\hat{C}\hat Y$.)
For these operatorial multiplications, $\hat Y$ was a two-derivative operator, whose trace of the logarithm
is known and can be easily verified. (This  was also checked independently below.) We emphasize that in the first two methods the resulting operators ($\hat{Y}\!\hat{C}$ and $\hat{C} \hat{Y}$ respectively) were six-derivative ones, while in the last one with double multiplication from both sides, $\hat Y\!\hat{C}\hat Y$  was an eight-derivative matrix-valued differential operator. At the end, all three described above methods of computation of ${\rm Tr}\ln \hat C$ agree for terms quadratic
in curvatures. These terms are only important for us here since they appear in the form of UV divergences of the theory (and are composed from  GR-invariants: $R^2$, $R_{\mu\nu}^2$, and $R_{\mu\nu\!\rho\sigma}^2$).

\item Similarly, the computation of ${\rm Tr}\ln \hat{Y}$ for the two-derivative operator  $\hat Y$  was verified using three analogous methods.  We used multiplication from  both sides by the operator $\hat A$
and also the symmetric form of multiplication  $\hat A\hat Y\!\!\hat A$,
where $\hat Y$ is a two-derivative operator, whose functional trace of the logarithm we searched here. Above, $\hat A$ was another two-derivative non-minimal spin-one vector gauge (massless) operator, whose
trace of the logarithm is well known \cite{Fradkin:1981iu} and can be easily found. Again, for the final results for ${\rm Tr}\ln \hat{Y}$ all three methods presented here agree to the order of  terms quadratic in curvatures $\cal R$.

\item In total divergent part $\Gamma_{\rm div}$ of the quantum effective action, we checked a complete cancellation  of terms with poles in $y=2\omega_C-3\omega_R$ variable, namely all terms with $\frac{1}{y}$ and $\frac{1}{y^{2}}$ in
denominators (originating from the expression for the gauge-fixing parameter $\beta$ in (\ref{mingauge})) completely cancel out. This is not a trivial cancellation
between the results of the following traces: ${\rm Tr}\ln \hat H$ and ${\rm Tr}\ln \hat M$.

\item Finally, using the same code written in Mathematica \cite{Wolfram} a similar computation in  four-derivative gravitational theory (Stelle theory in four dimensions) was repeated. We easily were able to reproduce all results about one-loop UV divergences there \cite{Avramidi:1985ki}.
To our satisfaction, we found a complete agreement for all the coefficients and the same non-trivial dependence on the parameter $x_{\rm 4-der}$, which was already defined above for Stelle gravity. This was the final check.

\end{enumerate}


\section{Structure of beta functions in six-derivative quantum gravity}
\label{s4}

\subsection{Limiting cases}
\label{s4.1}

In this subsection, we discuss various limiting cases of higher-derivative gravitational theories (both with four and six derivatives). We study in detail the situation when some of the coefficients of action terms in the Weyl basis tend to zero. We comment whether in such cases our method of computation is still valid and whether the final results for UV divergences are correct in that cases and whether they could be obtained by continuous limit procedures.

First, we discuss the situation with a possible degeneracy of the kinetic operator of the theory acting between quantum metric fluctuations $h_{\mu\nu}$ on the level of the quadratized action.
If the action of a theory in the UV regime has the following UV-leading terms
\begin{equation}
S_{{\rm grav}}=\!\int\! d^{d}x\sqrt{|g|}\left(\omega_{C,N}C\square^{N}C+\omega_{R,N}R\square^{N}R\right),
\end{equation}
with $\omega_{C,N}\neq0$ and also $\omega_{R,N}\neq0$
and after adding the proper gauge-fixing functional, then the kinetic operator can be defined, so in these circumstances it is not degenerate. Then it constitutes the operatorial kernel of the part of the action which is quadratic in the fluctuation fields.  It can be well-defined not only for the cases when $\omega_{C,N}\neq0$ and $\omega_{R,N}\neq0$, but also when $\omega_{R,N}=0$. This last assertion one can check by explicit inspection, but due to the length of the resulting expression we decided not to include such a bulky formula here. However, in the case $\omega_{C,N}=0$, a special procedure must be used to define the theory of perturbations and to extract UV divergences of the model. We remark that in the last case the theory is non-renormalizable. We also emphasize that the addition of the gauge-fixing functional here is necessary since without it the kinetic operator  (Hessian) is automatically degenerate as the result of gauge invariance of the theory (here in gravitational setup represented by the diffeomorphism gauge symmetry).

In general, as emphasized in \cite{Rachwal:2021bgb}, in four spacetime dimensions, the general UV divergences depend only on the coefficients appearing in the following UV-leading  part of the gravitational HD action,
\begin{multline}
S_{{\rm grav}}  = \! \int\! d^{4}x\sqrt{|g|}\left(\omega_{C,N}C\square^{N}C+\omega_{R,N}R\square^{N}R\right.\\
 +  \omega_{C,N-1}C\square^{N-1}C+\omega_{R,N-1}R\square^{N-1}R\\
\left.  +  \omega_{C,N-2}C\square^{N-2}C+\omega_{R,N-2}R\square^{N-2}R\right),
\label{eq: uvpart}
\end{multline}
where the last two lines contain subleading terms in the UV regime. However, they are the most relevant for the divergences proportional to the Ricci curvature scalar and also to the cosmological constant term \cite{Modesto:2017hzl}. Below for notational convenience, we adopt  the following convention specially suited for six-derivative gravitational theories, so in the case when $N=1$. We will call coupling coefficients in front of the leading terms as respective omega coefficients (like $\omega_C = \omega_{C,1}$ and analogously $\omega_R = \omega_{R,1}$), while the coefficients of the subleading terms with four derivatives we will denote as theta coefficients  (like $\theta_C = \omega_{C,0}$ and analogously $\theta_R = \omega_{R,0}$). Eventually, for the most subleading terms with subindex values of $(N-2)$ equal formally to $-1$ here, we have just one term contributing to the cosmological constant type of UV divergence. We denote this coefficient as $\omega_{-1}=\omega_{R,-1}$ and it is in front of the Ricci scalar term in the original classical action of the theory (\ref{lagr}). Simply this coefficient $\omega_{-1}$ is related to the value of the 4-dimensional gravitational Newton's constant $G_N$.

The expressions for the RG running of the cosmological constant and the Newton's constant  in \cite{Modesto:2017hzl,Rachwal:2021bgb} contain various fractions of parameters of the theory appearing in the action (\ref{eq: uvpart}). Still, for a generic value of the integer $N$, giving roughly the half of the order of higher derivatives in the model, we have the following schematic structure of these fractions:
\begin{equation}
\frac{\omega_{R,N-1}}{\omega_{R,N}}\, ,  \quad \frac{\omega_{C,N-1}}{\omega_{C,N}} \, , \quad \frac{\omega_{R,N-2}}{\omega_{R,N}} \, , \quad \frac{\omega_{C,N-2}}{\omega_{C,N}} \, .
\label{fractions}
\end{equation}
The structure of the UV divergences and of these fractions can be easily understood from the energy dimensionality arguments. We notice that in the Weyl basis with terms in (\ref{eq: uvpart}) written with Weyl tensors $C_{\mu\nu\!\rho\sigma}$ and Ricci scalars $R$ the only fractions, which appear in such subleading UV divergences are ``diagonal'' and do not mix terms from the spin-2 (Weyl) sector with terms from the spin-0 (Ricci scalar) sector. If in any of the above fractions, we take the limits: $\omega_{R,N-1}\to0$, $\omega_{C,N-1}\to0$,
$\omega_{R,N-2}\to0$, or $\omega_{C,N-2}\to0$, then the corresponding fractions and also related UV divergences (and resulting beta functions in question) simply vanish, provided that the
coefficients in their denominators $\omega_{R,N}$ and $\omega_{C,N}$ are non-zero.

On the other hand, if $\omega_{C,N}=0$, then we cannot rely on this limiting procedure.
In this case, on the level of the quadratized action the operator between quantum fluctuations is degenerate even after adding the gauge-fixing terms. This means that in this situation a special procedure has to be used to extract the UV divergences of the model. This is possible, but we will not discuss it here.

It is worth to notice that in turn, if $\omega_{R,N}=0$, then
the kinetic operator for small fluctuations and after adding the gauge fixing is still well-defined, as emphasized also above. For this case a special additional kind of gauge fixing has to be used, which fixes the value also of the trace of the metric fluctuations $h=g^{\mu\nu}h_{\mu\nu}$ and this last one in the theory with $N=0$  resembles the conformal gauge fixing of the trace.

If  $\omega_{R,N}=0$ and additionally $\omega_{R,N-1}$
or $\omega_{R,N-2}$ are non-zero, then the corresponding beta functions for the cosmological constant and Newton's gravitational constant
are indeed infinite and ill-defined as viewed naively from the expressions in (\ref{fractions}). This situation could be understood as that there is an additional new divergence not absorbed in the adopted renormalization scheme and the renormalizability of such a  theory is likely lost.
But if the model is with $\omega_{R,N}=0$ and at the same time $\omega_{R,N-1}=\omega_{R,N-2}=0$, then the contributions of corresponding fractions in (\ref{fractions})
are vanishing, because the limits $\omega_{R,N-1}\to0$ or $\omega_{R,N-2}\to0$
 must be taken as the first respectively. Only after this, the final limiting procedure $\omega_{R,N}\to0$
should be performed. Therefore, in this limiting situation, the proper
sequence of limits on respective fractions is as follows:
\begin{equation}
\lim_{\omega_{R,N}\to0}\left(\lim_{\omega_{R,N-1}\to0}\frac{\omega_{R,N-1}}{\omega_{R,N}}\right)=0
\end{equation}
 and
\begin{equation}
\lim_{\omega_{R,N}\to0}\left(\lim_{\omega_{R,N-2}\to0}\frac{\omega_{R,N-2}}{\omega_{R,N}}\right)=0\,.
\end{equation}
In this case, there are no contributions to the beta functions from these fractions, so the $R^2$ sector does not contribute anything to the mentioned UV divergences, while it is expected that the terms in the $C^2$ sector make some impact on beta functions.

However, the similar procedure cannot be applied in the sector with Weyl square terms ($C^2$ sector), so to the model with $\omega_{C,N}=0$ and at the same time $\omega_{C,N-1}=\omega_{C,N-2}=0$ since these cases have to be treated specially and separately. In the last case, after the limit, only the pure sector with Ricci scalar square terms ($R^2$ sector) survives and the theory is likely non-renormalizable. Then we expect contributions to UV divergences only from terms in the $R^2$ sector.

Regarding the divergences proportional to expressions quadratic in
curvatures ($R^2$, $C^2$, and the Gauss-Bonnet term ${\rm GB}$), we have found the following generic structure in four-derivative gravity \cite{Avramidi:1985ki}:
\begin{equation}
\frac{A_{-2}}{x^{2}_{\rm 4-der}}+\frac{A_{-1}}{x^{}_{\rm 4-der}}+A_{0},\label{eq: quadraticdivs}
\end{equation}
where in this case of four-derivative gravity the fundamental ratio of the theory is defined as
\begin{equation}
x_{\rm 4-der}=\frac{\omega_{C,0}}{\omega_{R,0}}=\frac{\theta_{C}}{\theta_{R}}.
\label{xdef4}
\end{equation}
The numerical coefficients $A_{-2}$, $A_{-1}$ and $A_{-0}$ are different for different types of UV divergences (here they are given by terms with four derivatives, namely by $R^2$, $C^2$ and $\rm GB$ terms respectively). The explicit numerical values are given in the formula (\ref{fourdergamma}).
One observes negative powers of the ratio $x_{\rm 4-der}$ in (\ref{eq: quadraticdivs}) and in (\ref{fourdergamma}), implying also negative powers of the coupling $\theta_C$ in the final results for these UV divergences.
This result signifies that the theory with $\theta_{C}=0$ should
be treated separately and then we do not have well-defined kinetic operator
 in a standard scheme of computation. The naive results with the
limit $\theta_{C}\to0$ of the above formula in (\ref{eq: quadraticdivs}) do not exist. Such theories with $\theta_C=0$ entail complete absence of gravitational terms in the $C^2$ sector. They are again very special and perturbatively non-renormalizable models. The above remarks apply both to pure $R^2$ Starobinsky theory as well as to theories in the $R^2$ sector with addition of the Einstein-Hilbert $R$ or the cosmological constant $\omega_\Lambda$ terms.

On the other side, the limit $\theta_{R}\to0$ in pure $C^2$ gravity seems not to produce any
problem with the degeneracy of the kinetic operator, nor with the final expression (\ref{eq: quadraticdivs}). The naive
answer would be just $A_{0}$ for (\ref{eq: quadraticdivs}) for each of the UV divergences in this case. But this is an \emph{incorrect} answer since for
pure four-derivative gravity with $\theta_{R}=0$ in the Weyl basis of terms, we have an enhancement of the symmetry in the model, beyond the case where $\theta_{R}$ was non-zero. In this situation, the theory enjoys also
conformal symmetry and a more specialized and delicate computation must be performed
to cover this case. This is the case of four-dimensional conformal (Weyl) gravity. (We decided for simplicity not to analyze  here the cases when besides the $C^2$ action for four-dimensional conformal gravity, there are also some subleading terms from the almost ``pure'' $R^2$ sector, that is $\omega_{-1}\neq0$ or when we allow for non-vanishing cosmological constant term $\omega_\Lambda\neq0$ -- these terms in the action would cause breaking of classical conformality.)

The computation in this case should reflect the fact
that also the conformal symmetry should be gauge-fixed. We remark that the conformal symmetry does not
require dynamical FP ghosts, because the conformal transformations of gravitational gauge
potentials (not the conformal Weyl gauge potentials $b_{\mu}$) are without
derivatives. At the end, when the more sophisticated method is employed, the eventual result is different than $A_{0}$ for each type out of three types of four-derivative UV-divergent GR-invariant terms in the quantum effective action of the model.
The strict result $A_0$ is still correct only for theories in which conformality is violated by inclusion of other non-conformal terms like the
Einstein-Hilbert $R$ term or the cosmological constant $\omega_\Lambda$ term.
We conclude that in the four-derivative theory, the two possible extreme cases of $\theta_{R}=0$ or $\theta_{C}=0$
are not covered by the general formula (\ref{eq: quadraticdivs}).  But in  each of these cases
the reasons for this omission  are different. In both these cases the separate more adapted methods of computation of UV divergences have to be used.

 In the case of six-derivative theory studied in \cite{Rachwal:2021bgb}, we have the following structure of UV divergences quadratic in gravitational curvatures
\begin{equation}
B_{0}+B_{1}x,\label{eq: sexticdivs}
\end{equation}
with new values for the constants $B_{0}$ and $B_{1}$. The explicit numerical values are given in our formula (\ref{sixdergamma}) with the results. We also remark that the values of the constant terms $B_{0}$ are different than the values of $A_{0}$ in the previous four-derivative gravity case. Moreover, the numerical coefficients $B_{0}$ and $B_{1}$  are different for different types of UV divergences of the effective action ($R^2$, $C^2$ and $\rm GB$ terms respectively).
When the leading dynamics in the UV regime is governed by the theory with six derivatives, then the fundamental ratio $x$ we define as
\begin{equation}
x=\frac{\omega_{C,1}}{\omega_{R,1}}=\frac{\omega_{C}}{\omega_{R}}.
\label{xdef6}
\end{equation}
 We emphasize that in such a case, we cannot continuously take the limit $\omega_{C}\to0$. Although,
naively this would mean the limit $x\to0$, the result just $B_{0}$ from (\ref{eq: sexticdivs}) would be incorrect.
This is because in this case we cannot trust the method of the computation. When
$\omega_{C}=0$ the kinetic operator is degenerate (the same as it was in the four-derivative gravity case)
and needs non-standard
treatment, that we will not discuss here.

 Moreover,  looking at the last formula (\ref{eq: sexticdivs}), the other limit $\omega_{R}\to0$ is clearly
impossible too, because it gives divergent
results. However, in this case ($\omega_{R}=0$) and on the contrary to the previous case ($\omega_{C}=0$), we could trust the computation at
least on the level of the kinetic operator (Hessian) and its subsequent computation of the functional trace of the logarithm of.  In this case, the final divergent results in (\ref{eq: sexticdivs}) signify
that the theory likely is  non-renormalizable and that there are new UV divergences
besides those ones derived from naive power counting analysis\footnote{We remark that the generic power counting analysis of UV divergences in six-derivative quantum gravity, as presented in section \ref{s1.1}, applies only in cases when $\omega_C\neq0$ and $\omega_R\neq0$.}. We conclude that in the case of six-derivative gravity, both cases
$\omega_{R}=0$ or $\omega_{C}=0$  require special treatment and the type of formula like in  (\ref{eq: sexticdivs}) or (\ref{sixdergamma}) does not apply there and the limiting cases are not continuous. More discussion of these limits is contained also in the further subsection \ref{s4.4}.


\subsection{Dependence of the final results on the fundamental ratio $x$}
\label{s4.2}

Here we just want to understand
the $x$-dependence in the result for the beta functions in  six-derivative gravitational theory. We first try to
analyze the situation for simpler theory (with four derivatives), prepare the ground for the theory with six derivatives, and then eventually draw some comparison between the two. We look for singular $\frac{1}{\omega_R}$ or $\frac{1}{\omega_C}$ dependence (corresponding to positive or negative powers of the fundamental ratio $x=\frac{\omega_C}{\omega_R}$ respectively) in functional traces of the fundamental operators defining the dynamics of quantum perturbations important to the one-loop perturbative level. We note that the two definitions for the ratio $x$ in (\ref{xdef4}) and in (\ref{xdef6}) respectively for four- and six-derivative gravities are compatible with each other and the proper use of them (with theta or omega couplings) is obvious in the specific contexts they are used in. Below, when we will refer to features shared by both four- and six-derivative gravitational theories, we will use common notation with general $\omega_C$, $\omega_R$ and $x$ coefficients and we will not distinguish and not change it to the special notation originally adequate only to Stelle quadratic theory (with $\theta_C$, $\theta_R$ and $x_{\rm 4-der}$). We hope that this will not lead to any confusion.

We emphasize, that when we have one of the two terms missing -- with $\omega_R$ or $\omega_C$ front couplings -- in the leading in UV part of the action of the model, then the theory is badly non-renormalizable and degenerate. For example, one cannot define even at the tree-level the flat spacetime graviton propagator since the parts proportional to $P^{(0)}$ or $P^{(2)}$ projectors do not exist in cases when $\omega_R=0$ or $\omega_C=0$ respectively. However, there we can still use the Barvinsky-Vilkovisky (BV) trace technology to compute the new UV divergences. The fact that they are not possible to be absorbed in counterterms of the original theory is another story related to the non-renormalizability of the model that we will not discuss further here. We think that, for example, using the BV technique one can fast compute UV divergences in Einstein-Hilbert (E-H) theory in $d=4$ (which is a non-renormalizable model) and this method still gives a definite result (besides that these divergences are gauge-fixing dependent and valid only for one gauge choice). Moreover, using the BV traces machinery and the minimal form of the kinetic operator is essential to get final
results for the unique effective action (as introduced by Barvinsky \cite{Barvinsky:1985an,Barvinsky:1990up}), also in perturbatively non-renormalizable models.

In quadratic gravity (four-derivative theory) in $d=4$, setting $\theta_C=0$ is highly problematic. The same regards taking the limit $\theta_C\to0$, because then the pure $R^2$ theory can be fully gauge-fixed. And for example, this means that on flat spacetime background, the kinetic operator vanishes, perturbative modes are not dynamical and there is no graviton propagator. Using the standard technique of the one-loop effective action one sees that the traces of the functional logarithms of $\hat H$ and of $\hat C$ operators
both contain singular expressions $\frac{1}{\theta_C}$, and there is no final cancellation between them. In this case of four-derivative gravity, in final results for UV divergences,  we really see inverse powers of the fundamental ratio of the theory $x_{\rm 4-der}$.

The results in quadratic Stelle gravity, when we set $\theta_R=0$, are not continuous either. Because in this case
the  local gauge symmetry of the theory is enhanced. We have also conformal symmetry there. The model is identical to the Weyl gravity in $d=4$ described by the action $C^2$.
As emphasized in \cite{Fradkin:1981iu}, 
 this case of $\theta_R=0$ has to be treated specially. Also, in this model, the
conformal symmetry has to be gauge-fixed and in this special case the operators $\hat H$ and $\hat C$ are different than their
limiting versions under $\theta_R\to0$ limit from the generic four-derivative theory case.  Hence also the results for the beta functions are different than the limits of the corresponding beta functions in the situation with $\theta_R \neq0$.

If we start with the theory with $\theta_C=0$ from the beginning, then there are serious problems with the kinetic operator.  We checked that it cannot be put by standard fixing of the gauge to the
minimal form with four-derivative leading operator. Moreover, as the result of this process one of the typical  gauge-fixing parameters remains undetermined. Here one can try to compute the trace of the  logarithm of the Hessian using
the method proposed in \cite{Modesto:2017hzl} 
consisting of multiplying by some two-derivative non-minimal operator and getting
a six-derivative operator, whose trace can be easily found. But it is hard to believe that one has any chance to get a non-singular answer for all the beta functions in pure $R^2$ theory since it is known that this theory is non-renormalizable (because it lacks the $C^2$ counterterm in the bare action).

Actually, here (for the $\theta_C=0$ case) one could choose the $\hat C$ matrix-valued differential operator different from the standard minimal prescription and choose different
values for the $\gamma$ gauge-fixing parameter. In the standard minimal choice for the gauge-fixing parameters and in this model, the $\hat C$ matrix contains an
irregular part in $\theta_C$ coupling ($\frac{1}{\theta_C}$ pole), because of the dependence of $\gamma$ on $\theta_C$. This last dependence originates from the conditions forced on gauge-fixing parameters in order to put the kinetic operator in the minimal form, in the standard case $\theta_C\neq0$. However, knowing that in the case with $\theta_C=0$, this procedure is anyhow unsuccessful, we have the freedom  to choose the value of $\gamma$ different than the standard one and at our wish.

In principle, similar considerations can be repeated verbatim for the case of six-derivative theory (with $N=1$ power
exponent on the box operator in the defining the theory action in (\ref{eq: uvpart})). But we remark here that the theory with $\omega_R=0$ and $N=1$ is \emph{not} conformally invariant in  $d=4$ dimensions. And the above problems
with the gauge fixing of the Hessian operator $\hat H$ and non-minimality of it  in the case $\omega_C=0$ still persist. This is because here  for six-derivative gravitational theories the box operator $\square$ acting between two gravitational curvatures is only a spectator from the point of view of the UV-leading part of the $\hat H$ operator (with the highest number of derivatives and with the zeroth powers in gravitational curvatures) or from the point of view of
flat spacetime kinetic operator and flat spacetime graviton's propagator. The box operator in momentum space gives only one additional factor of $-k^2$ to the kinetic operator and additional suppression by $-k^{-2}$ to the propagator.
The Hessian in the six-derivative theory with $\omega_R=0$  must possess the same definitional  issues as the one in the four-derivative theory (with $\theta_R=0$), because for the kinetic terms box operator again plays only the role of the spectator. Hence the difference on this level between four- and six-derivative theories is only in some overall multiplicative coefficient (like flat spacetime d'Alembertian operator $\partial^2$ is $-k^2$ in Fourier space). So then, if we know that the Hessian $\hat H$ is almost well-defined  for the conformal gravity case (up to the need for additional gauge fixing of the conformal symmetry), then the same will be true for the Hessian in the six-derivative theory with the $\omega_R=0$ condition in $d=4$ spacetime dimensions, although then the theory ceases to be conformal anymore. In conformal gravity in $d=4$, when $\theta_R=0$,  we have almost well-defined Hessian, because  we know that it gives rise to a good renormalizable theory at least to the one-loop perturbative level of computations.

 Now, also in the case of six-derivative theories, setting $\omega_R=0$ does not create any problem for the form of neither $\hat H$ nor  $\hat C$ operators. Only the final results for the beta functions show $\frac{1}{\omega_R}$ poles as this was manifest from the results in \cite{Rachwal:2021bgb}. 
In turn, in six-derivative theories, the limit $\omega_C\to0$ seems regular, but it is questionable that now we can trust the results of this limit. In the  pure $ R\square R$ theory, we expect to get some discontinuous results for the beta functions not obtainable by the limit $\omega_C\to0$ since this model is non-renormalizable. In this model, there is still an open problem that  one cannot make the kinetic operator of fluctuations a minimal 6-derivative one. Furthermore, taking the limit  $\omega_C\to0$ on the kinetic operator from the generic case $\omega_C\neq0$  produces a Hessian $\hat H$ that vanishes on flat spacetime. Hence it seems that in this case the intermediate steps of the process of computing the divergent part of the effective action are not well-defined, while the final result is amenable to taking the limit $\omega_C\to0$, but exactly because of this former reason, we should not trust these apparently continuously looking limits.

One should analyze deeper the form of the leading in the number of derivatives (and also in the UV regime) part of the kinetic operator $\hat H $ of the theory between graviton fluctuations. The insertions of box operators, like any power or functions of the box operator $\square$, are  only the immaterial differences between the cases of four- and six-derivative theories here. These operators are only spectators for getting the leading part of the Hessian, which is with the highest number of derivatives and also considered on flat spacetime, so with the condition that ${\cal R}=0$. Using formula (\ref{bilin}) with solutions for gauge-fixing parameters as in (\ref{mingauge}), one finds in the generic case $\omega_C\neq0$ and $\omega_R\neq0$, that the kinetic operator (leading part of the Hessian) is indeed minimal and of the form
\begin{multline}
H_{{\rm lead}}^{\mu\nu,\rho\sigma}=\frac{\omega_{C}}{2}\square\left(g^{\mu\rho}g^{\nu\sigma}+g^{\mu\sigma}g^{\nu\rho}\right)\\
-\omega_{C}\frac{\omega_{C}-6\omega_{R}}{4\omega_{C}-6\omega_{R}}\square g^{\mu\nu}g^{\rho\sigma}\,.
\label{hessianred}
\end{multline}
In the above formula, one does not see any singularity when $\omega_C$ is vanishing (one saw $\omega_C^{-1}$ divergences in the expressions for $\alpha$ and $\gamma$ parameters in (\ref{mingauge})), but in this case the above treatment was not justified. When $\omega_C=0$, one can solve the system for gauge-fixing parameters for $\beta$  and $\gamma'=\frac{\gamma}{\alpha}$ and assume that formally $\frac{1}{\alpha}=0$ and $\frac{1}{\gamma}=0$, but in the ratio $\frac{\gamma}{\alpha}$ the limit is finite. One then finds that $\beta=1$ and $\gamma'=-2\omega_{R}$ and after substitution to the original Hessian, one gets that its leading part explicitly vanishes. The same one gets by plugging the naive limit $\omega_C\to0$ in (\ref{hessianred}). One also sees from the explicit solutions in (\ref{mingauge}) and resulting general expression for $\gamma'$ (i.e. $\gamma'=\frac{4}{3}\omega_{C}-2\omega_{R}$) that by plugging $\omega_C=0$ one finds again that $\beta=1$ and $\gamma'=-2\omega_R$ as derived exactly above. The highest derivative level of the gravitational action is then completely gauge-fixed.

In the opposite case, when $\omega_R=0$, the leading part of the Hessian does not vanish, but it is degenerate and in the form
\begin{equation}
H_{{\rm lead}}^{\mu\nu,\rho\sigma}=\frac{\omega_{C}}{2}\square\left(g^{\mu\rho}g^{\nu\sigma}+g^{\mu\sigma}g^{\nu\rho}\right)-\frac{\omega_{C}}{4}\square g^{\mu\nu}g^{\rho\sigma},
\end{equation}
because this operator does not possess a well-defined inverse, precisely in $d=4$ dimensions. An addition of a new conformal-like type of gauge-fixing here $\tau h\square^{3}h$  with a new (fourth) gauge-fixing parameter $\tau$ and where the trace of metric fluctuations $h=h^\mu{}_\mu$ is used, removes the degeneracy provided that $\tau\neq0$ is selected. Then the kinetic operator takes the form
\begin{multline}
H_{{\rm lead}}^{\mu\nu,\rho\sigma}=\frac{\omega_{C}}{2}\square\left(g^{\mu\rho}g^{\nu\sigma}+g^{\mu\sigma}g^{\nu\rho}\right)\\
+\left(\tau-\frac{\omega_{C}}{4}\right)\square g^{\mu\nu}g^{\rho\sigma}.
\end{multline}
Moreover, for any non-zero value of $\tau$ the Hessian is still a minimal operator. For $\tau\neq0$  the inverse exists and also the propagator can be defined around flat spacetime. The only question is whether the final results are $\tau$-independent since this is a spurious gauge-fixing parameter. The reason for such independence is obvious in the four-derivative case, since $\tau$  is a gauge-fixing parameter for conformal symmetry (conformal gauge-fixing parameter, so this is then in such circumstances a symmetry argument). But in the case of six-derivative model in $d=4$, the reasoning with conformal symmetry is not adequate since this model is not conformal anymore. Only the explicit computation may show that $\tau$ parameter drops out from final results as it should for them to be physical and $\tau$  gauge choice independent.

In four-derivative gravitational theory, one can see the dependence on the $x_{\rm 4-der}$ ratio only in the coefficient of the $R^2$ counterterm. This dependence is with the general schematic form  $A_{-2}x_{\rm4-der}^{-2} +A_{-1} x_{\rm4-der}^{-1} +A_{0} x_{\rm4-der}^0$ like in (\ref{eq: quadraticdivs}) and in (\ref{fourdergamma}). We remark that for other counterterms (namely for $C^2$ and ${\rm GB}$ in this Weyl basis), the coefficients of UV divergences are numbers completely independent of $x_{\rm4-der}$. One could try to explain here this quadratic dependence in the inverse ratio  $x_{\rm4-der}^{-1}$ in front of the $R^2$ counterterm in a spirit similar to the argumentation presented in \cite{Modesto:2017hzl}, where we counted active degrees of freedom contributing to the corresponding beta functions of the theory. It is well known by the examples of beta functions in QED coupled to some charged matter and in Yang-Mills theory, that the beta function at the one-loop level expresses weighted counting of degrees of freedom and their charges in interactions  with gauge bosons in question (minimal couplings in three-leg vertices are enough to be considered here due the gauge symmetry). The similar counting could be attempted here, but in gravity, especially in HD gravity, there is a plenty of other gravitational degrees of freedom, so this is quite a difficult task to enumerate all of them and their strength of interactions  in cubic vertices when they interact with background gravitational potentials. Therefore, this task of explaining $x$-dependence and numbers present in the expressions for all the beta functions both in four- and six-derivative theories,  now seems to be too ambitious and we leave it for some further future considerations.

Instead, we comment briefly on the general dependence on the $x_{\rm4-der}$ ratio in four-derivative theory and compare this with six-derivative theory. In the case of $N=1$ (six-derivative gravitational theory), it was  found as a main result in \cite{Rachwal:2021bgb}, that the dependence on $x$ is only in front of the $C^2$ counterterm and this is a linear dependence $B_1x^1+B_0 x^0$ like in (\ref{eq: sexticdivs}) with non-negative powers of the $x$ ratio. The other counterterms $R^2$ and ${\rm GB}$ are with constant coefficients (only $B_0$ terms present) (cf. with (\ref{sixdergamma})). If the other than the Weyl basis is used to write counterterms, then the $x$-dependence is linear in coefficients in front of each basis term (like in the basis  with  $R^2$, $R^2_{\mu\nu}$, and $R^2_{\mu\nu\rho\sigma}$ terms). These explicit dissimilarities between $N=0$ and $N=1$ models certainly require deeper investigations.

It is interesting also to analyze a special value of the fundamental ratio $x$ of the six-derivative gravitational theory, which makes the $C^2$ sector of UV divergences completely finite. This value is exactly $x=-\frac{3573}{80}=-44.6625$. The $R^2$ sector of UV divergences cannot be made finite this way. We remind for comparison that in the case of quadratic gravity with four derivatives in $d=4$, the special values for $x_{\rm4-der}$, which made \emph{contrary} the $R^2$ sector UV-finite, were two and they were $x_{\rm4-der}=3 (3 \pm \sqrt {7})$ (their numerical approximations are as follows: $x_{\rm4-der,}{}_-\approx1.0627$ and $x_{\rm4-der,}{}_+\approx16.937$) as solutions of some non-degenerate quadratic algebraic equation. Again, contrary to the previous case with six derivatives, here the divergences in the $C^2$ sector cannot be made vanish.

 Now, we discuss the differences between the two extreme cases $\omega_C=0$ and $\omega_R=0$.
In six-derivative model or when we have even more derivatives, superficially these two couplings and their roles for the computation of UV divergences may look symmetric. This is however not true due to the different impact of these two conditions on the form of the kinetic operator $\hat H$. In the case when $\omega_R=0$, the Hessian operator still exists, while for $\omega_C=0$ we lose its form. This observation has profound implications as we explain below. First, it is a fact that both these conditions lead to badly non-renormalizable theories, in which the flat spacetime propagator cannot be simply defined. Moreover, if $N>0$ in none of these two reduced models we have an enhancement of symmetries and none of them has anything to do with conformal gravity models, which are present only for $N=0$ and $\omega_R=0$  case, despite that in constructions of these six-derivative models we might use only terms with Weyl tensor. (However, here we use the term $C\square C$, where it is known that the GR-covariant box operator $\square$ is not conformal.)

Our explanation of the $x$-dependence is as follows. First, in the generic model with $N>0$, since it happens that it is the $N=0$ scale-invariant gravitational model which is here the exceptional one. For six-derivative theory (or any one with $N>0$) the two reduced models with conditions that $\omega_C=0$ or $\omega_R=0$ respectively are not renormalizable and likely even at the one-loop level higher types of divergences (besides $C^2$ and $R^2$ from (\ref{sdiv})) will be generated. From this we expect that there must be some problems with UV divergences obtained from naive power counting arguments here. The problems must show up somehow in the final numerical values for divergent terms or in the intermediate steps of the process of computing these divergences. These problems then signal that we are working with non-renormalizable theory, which do not have a good control over perturbative divergences showing up at the one-loop level.

First, in the case $\omega_C=0$, we see that the problems are already there with the definition of the kinetic operator (Hessian) between quantum metric fluctuation fields. This implies that further processing with this operator is ill-defined, we cannot trust it and even if it would give us some final results for divergences, then they are not reliable at all since the theory is non-renormalizable. But we already found here the instance of the problem, which makes our final limiting results (in the $\omega_C\to0$ limit) not trustable.
This means that from the expression in (\ref{eq: sexticdivs}), we do not expect any additional obstacles, like
$\frac{1}{\omega_C}$ poles since the price for non-renormalizability was already paid and we have already
met dangerous problems, which signal the incorrectness of the naive limit $\omega_C\to0$. This should already take away our trust in the limit $\omega_C\to0$ of expressions for UV divergences in (\ref{sixdergamma}). Then this line of thought in the case $\omega_C=0$ does not put any constraints at all on the final form of the $x$-dependence in (\ref{eq: sexticdivs}) since these results like in (\ref{eq: sexticdivs}) in the limit $\omega_C\to0$ will anyway be likely incorrect.

Now, in the other case with $N>0$ and $\omega_R=0$, we do not have the problem with the definition of the Hessian ${\hat H}$. Formally, we can process it till the end of taking the functional trace of the logarithm and adding contributions from ${\rm Tr}\ln \hat M$  and  ${\rm Tr}\ln \hat C$. But somehow, we must find the occurrence of the problem, because the theory is non-renormalizable! So the only place in which the problem may sit is in the final $x$-dependence of the results for UV divergences. These results should be ill-defined, when the limit $\omega_R\to0$ is attempted. And this implies that they must be with poles in the $\omega_R$ coefficient, so they must be instead with positive powers of the $x$ ratio of the theory. Hence, we conclude that the $x$-dependence must be linear or quadratic, but always with positive powers of the ratio $x$. This is now confirmed by explicit form as in (\ref{sixdergamma}) for UV divergences of six-derivative theory. The problems with renormalizability of the pure theory $C\square C$ show up in the last possible moment in the procedure for obtaining the result, namely when one wants to take the limit $\omega_R\to0$ or equivalently $x\to\infty$. This is the generic situation for any super-renormalizable theory and for any $N>0$.
There are still some mysterious things here, like why the dependence is only linear in $x$ and why only for the $C^2$ type of UV counterterm, while two other counterterms $R^2$ and $\rm GB$ are numbers completely independent of $x$. Right now we cannot provide satisfactory mathematical explanations for these facts.

Using this argumentation in the theory models with $N=1$, we get an explanation for the $x$-dependence in
formula for UV divergences in (\ref{eq: sexticdivs}). The logical chain for the explanation should be as
follows. Firstly, in the pure theory $C\square C$, one concludes that the problems of non-renormalizability
shows only in the final results as impossibility to take the limit $\omega_R\to0$ or equivalently
$x\to\infty$ of formula in (\ref{eq: sexticdivs}) for divergences of the model. Hence the dependence must
be on non-negative powers of the ratio $x$ in formula (\ref{eq: sexticdivs}), as it is clearly confirmed
by explicit inspection of this formula. This settles the issue of the structure of exact beta functions for $N=1$ models (and
also for higher $N\geqslant1$ cases too). Now, the same formula is a starting point for an attempt to
take the other limit $x\to0$ of the also non-renormalizable model of the type $R\square R$. But in such a
model we have already found a source of the problem caused by non-renormalizability earlier, that it is
here connected with the impossibility to properly define non-degenerate Hessian operator in the model. But this
limiting case of $x\to0$ must follow the same structure as already established  in formula (\ref{eq: sexticdivs}). Simply,
theoretically speaking, there is no need to see more instances of problems due to non-renormalizability in
the $R\square R$ model. Hence, the first explanation based on the $C\square C$ model is sufficient and the
results in the model $R \square R$ must be consistent with it. Moreover, from just the analysis of the
case $x\to\infty$, we have concluded what is the structure in a generic renormalizable case, when we have both
$\omega_R\neq0$ and $\omega_C\neq0$ (so $x\neq0$ and $x\neq\infty$). This structure is beautifully confirmed by the formula (\ref{eq:
sexticdivs}) or (\ref{sixdergamma}) explicitly for the generic case.

As emphasized above, it is in turn the $N=0$ case, which is extraordinary and it changes the pattern of $x_{\rm4-der}$-dependencies described above.
This all can be traced back to the fact that for $N=0$  we have the possibility of reducing the generic HD scale-invariant model to the conformal one, when the full conformal symmetry is preserved (at least on the classical level of the theory). This happens, when one takes the isolated case of $\theta_R=0$ and $\theta_C\neq0$ for the four-derivative theory (for positive-definiteness we may also assume that $\theta_C>0$). This case is discontinuous and cannot be taken as the naive limit $x_{\rm4-der}\to\infty$ of the formula (\ref{eq: quadraticdivs})  for the $R^2$ type of UV divergences, which would leave us effectively only with the $A_0$ coefficient. It is well known that the conformal gravity model is renormalizable one (at least to the one-loop level), contrary to the case of the theory $C\square C$, which was discussed above. This means that we shall not find any source of the problem when computing and getting results for UV divergences of this $C^2$ model. We do not find problems with the Hessian or the propagator provided we also gauge-fix the conformal local symmetry of Weyl conformal gravity. We shall not find the problem with the final expression of UV divergences, so there we shall not expect poles with $\theta_R$ coefficient. But some $x_{\rm4-der}$-dependence up to the quadratic order could be present (this is due to the one-loop character of the computation here; one can understand this easily from contributing Feynman diagrams). So then, we conclude that this dependence may be only in positive powers of the inverse ratio, namely of $x_{\rm4-der}^{-1} = \theta_R /\theta_C$. This is again confirmed in the formula (\ref{eq: quadraticdivs}) and (\ref{fourdergamma}), where we indeed find the quadratic dependence but in the inverse ratio $x_{\rm4-der}^{-1}$.

Simply, the final results for the generic case $x_{\rm4-der}\neq0$ and $x_{\rm4-der}\neq\infty$ cannot depend on positive powers of $x_{\rm4-der}$ since then the limit of conformal gravity in $d=4$ (i.e. $x_{\rm4-der}\to\infty$) would produce divergent results, but we know that Weyl gravity is renormalizable with a good control over one-loop UV divergences. However, this does not mean that the results for conformal gravity are continuous and obtainable from the generic ones in (\ref{fourdergamma}) by taking the limit $x_{\rm4-der}\to\infty$ there. We admit the fact that the coefficients there may show some finite discontinuities. However, both in the true and naive $x_{\rm4-der}\to\infty$ limiting forms they must be finite -- we only exclude the case when they would be divergent in the $x_{\rm4-der}\to\infty$ limit. In this way, the results in renormalizable 4-dimensional conformal gravity for UV divergences may be expressed via finite numbers multiplying just one common overall divergence (like $1/\epsilon$ parameter in dimensional regularization (DIMREG) scheme). The theory is renormalizable and there are no new divergences inside coefficients of established form of UV divergences in generic HD Stelle theory in $d=4$ spacetime dimensions, as in (\ref{sdiv}).

The significant difference between the cases of $N=0$ and $N\geqslant1$ is that in the former case the theory with $\theta_R=0$ is conformal on the classical tree-level as well as on the first quantum loop, since we know that Weyl conformal quantum gravity is one-loop renormalizable. This is why the pattern of the $x$-dependence in these two cases is diametrically different. In both these cases of $N=0$ and $N\neq0$, one can derive the general structure of beta functions in generic HD theory with any finite value of the fundamental ratio $x$ ($x\neq0$ and $x\neq\infty$) by just analyzing the limit $x\to\infty$ (or respectively $x_{\rm4-der}\to\infty$) and its divergences which should or should not appear there respectively for the cases of $N\neq0$  or $N=0$.

The inverse quadratic dependence on the ratio $x_{\rm4-der}$ in the case of four-derivative Stelle theory can be easily understood as well. It is up to the quadratic order and the same dependence we would expect in the case of six-derivative gravitational theory in the $C^2$ sector of UV divergences. However, there as a surprise we find only up to linear dependence on the fundamental ratio $x$ and only in one distinguished sector of $C^2$ divergences. In general, we can have up to quadratic dependence on $x$ in six-derivative models or on $x_{\rm4-der}^{-1}$ in the Stelle gravity case in $d=4$ spacetime dimensions. The UV divergences of some renormalizable HD gravity models in $d=4$ spacetime dimensions are all at most quadratic in the general gravitational curvature (schematically they are ${\cal R}^2$). Hence they can be all  read from the one-loop perturbative quantum corrections to the two-point graviton Green function, so equivalently from the quantum dressed graviton's propagator around flat spacetime background. We remind that here there is no quantum divergent renormalization of the graviton wave function. Moreover, higher orders in graviton fields (appearing in interaction vertices) are completely determined here due to the gauge invariance (diffeomorphism) present in any QG model, so we can concentrate below only on these two-point Green functions.

As it is known from diagrammatics, here the contributing Feynman diagrams may have either one propagator (topology of the bubble attached to the line) or two propagators (sunset diagrams) at the one-loop order and for corrections to the two-point function. In the most difficult case, there are here two perturbative propagators. Since in our higher-derivative theory we have two leading terms shaping the UV form of the graviton's propagator, namely the terms $\omega_C C\square C$ and $\omega_R R\square R$, then the corresponding propagator may be either with the front coefficient $\omega_C^{-1}$ or $\omega_R^{-1}$ respectively as the leading term. To change between the two expansions (in $\omega_C$ or in $\omega_R$) one needs to use one power of the ratio $x$. Since we have two such propagators in the one-loop diagrams considered here, then dependence is up to the quadratic power in $x$. Sometimes we need to change back from $\omega_C$  to $\omega_R$ as the leading coefficient of the tree-level propagator, and then we need to multiply by inverse powers of the ratio $x$. The quadratic dependence is what we can have here in the most complicated case, which is actually realized in Stelle generic theory with both $\theta_C\neq0$ and $\theta_R\neq0$. (The argumentation above can be repeated very similarly for quadratic gravity in $d=4$ forgetting about one power of box operator $\square$ and changing corresponding omega coefficients to theta coefficients and $x$ to $x_{\rm4-der}$.) Apparently, in the case of six-derivative gravitational theories there is some, for the moment, unexplained cancellation, and we see there only the dependence up to the first power on the ratio $x$ of that theory.

One should acknowledge here the speciality of the case of $d=4$ and one-loop type of computation. For
higher loop orders the powers of the $x$ ratio may appear higher in the final expressions for UV divergences
of the theory. Similarly, if one goes to higher dimensional QG models, then even in renormalizable models
at the one-loop level, one needs to compute higher $n$-point Green function. This is because in even
dimension $d$ one needs in renormalizable theory not only to renormalize terms of the type ${\cal
R}\square^{(d-4)/2}{\cal R}$ but also others with more curvatures (and correspondingly less powers of
covariant derivatives) down to the term of the type ${\cal R}^{d/2}$, where we do not find covariant
derivatives acting on curvature at all. In the middle, the general terms can be schematically parametrized
as $\nabla^{d-2i}{\cal R}^{i}$ for $i=2,\ldots,\frac{d}{2}$ -- all these terms have the energy
dimensionality equal to the dimensionality of spacetime $d$. For the last term of the type ${\cal
R}^{d/2}$  one needs to look at the quantum dressed $n=\frac{d}{2}$-point function at the one-loop order.
In conclusion, in higher dimensions one should consider not only two-point functions with one-loop
diagrams with the two topologies described above, but up to quantum dressed $\frac{d}{2}$-point functions.
And even for one-loop perturbative level these additional diagrams may have more complicated topology
meaning more vertices and more propagators and this means that also powers of the ratio $x$ or $x^{-1}$
respectively will be higher and higher. They are expected to be up to the upper bound given by the
maximal power exponent equal to $\frac{d}{2}$ -- this can be derived from the expression of quantum
dressed $\frac{d}{2}$-point function, which is built exactly with $\frac{d}{2}$ propagators joining
precisely $\frac{d}{2}$ 3-leg the same perturbative vertices. Then the topology of such  a diagram is this one of the
main one-loop ring and $\frac{d}{2}$ external legs attached to it, with each one separately and each one emanating
 from one single $3$-leg vertex. Again the situation at the one-loop and in $d=4$  is quite
special and simple since the ratio $x$ appears here only up to the maximal power exponent given by $\frac{d}{2}=2$.

As a side result, one also sees that the situation in four-derivative model with the condition $\theta_C=0$ is somehow ``doubly'' bad. First, the Hessian is not well-defined to start with and this takes away our trust in this type of computation. Moreover, if we would attempt to take the limit $\theta_C\to0$ (or equivalently $x_{\rm4-der}\to0$) in the final result like in (\ref{eq: quadraticdivs}), then we get a second problem since such limit gives infinite results. This means that we somehow doubly confirmed the problem with the perturbative and multiplicative renormalizability of such a model. It is not that the two instances of the problem support each other -- they appear somehow independently and are not related, nor they cancel out.  Above, we have seen that in the six-derivative (or general $N>0$) case, they could occur completely independently for two completely different types of non-renormalizable theories (with the conditions of $\omega_R=0$  or $\omega_C=0$ respectively). Here, we see that since conformal gravity at one-loop must be without any problem of this type (no problem with the Hessian and no problem with getting infinite results of the limits $x_{\rm4-der}\to\infty$), then the occurrence of these two problems at the same time must happen in badly non-renormalizable model with $\theta_C=0$  condition. In other words, since conformal gravity is a safe model, then the model with $\theta_C=0$ must suffer twice since all these two problems must inevitably appear in one model or the other, if some extreme special cases like $\theta_R=0$ or $\theta_C=0$ are being considered.

Again, we remark that in generic quadratic gravity model we see up to quadratic dependence on the inverse ratio  $x_{\rm4-der}^{-1}$, but a precise location where this dependence shows up is still not amenable for an easy explanation. We do not know why this happens in the $R^2$ sector only, while the $C^2$ and $\rm GB$ sectors are free from any $x_{\rm4-der}$-dependence. But at least the dependence on the inverse ratio $x_{\rm4-der}^{-1}$, rather than on its original form $x_{\rm4-der}=\frac{\theta_C}{\theta_R}$, in the exceptional case of $N=0$ can be explained by the miraculous one-loop perturbative renormalizability of the conformal gravity model in $d=4$.

\subsection{Case of Conformal Gravity}
\label{s4.3}

Here we continue the discussion of  related issues, but now in the framework of conformal gravity, so within the model with the reduced HD action  with $N=0$ and formally with $\theta_R=0$. There are various motivations for conformal gravity in $d=4$ spacetime dimensions \cite{Mannheim:2011ds,Mannheim:2016gwb}. As it is well known the reason for the multiplicative renormalizability of such a reduced model, when we have from the beginning that $\theta_R=0$ is the presence of conformality - conformal symmetry both on the tree level and also on the level of the first loop. Unfortunately, the story with conformal gravity in $d=4$ is even more complicated than what we argued above. First, already at the one-loop level one discovers the presence of conformal anomaly, which is typically thought as not so harmful on the first loop level. However, it heralds the soon breaking of conformal symmetry like for example via the appearance of the $R^2$ counterterm at the two-loop level. Such term as a counterterm is not fully invariant under local conformal transformations - it is only invariant under so called restricted conformal transformations that is such transformations whose parameters satisfy the source-free background GR-covariant d'Alembert equation ($\square\Omega=0$) on a general spacetime. Hence the $R^2$ term is still scale-invariant but it breaks full conformal symmetry of the quantum conformal gravity. It seems the only way out of this conformal anomaly problem is to include and couple to conformal gravity specific matter sector to cancel in effect the anomaly. This is, for example, done in ${\cal N}=4$ conformal supergravity coupled to two copies of ${\cal N}=4$ super-Yang-Mills theory, first proposed by Fradkin and Tseytlin \cite{Fradkin:1985am}. In such a coupled supergravity model, we have vanishing beta functions, implying complete UV-finiteness and conformality present also on the quantum level. This is conformal symmetry in the local version  (not a rigid one) with Weyl conformal transformations in the gravitational setup and on the quantum field theory level.

If not in the framework of ${\cal N}=4$ Fradkin and Tseytlin supergravity, the conformal anomaly of local conformal symmetry signals breaking of conformal symmetry, while scale-invariance (global part of it) still may remain intact. In the long run, besides the presence of non-conformal $R^2$ counterterm, this breaking will put conformal Ward identities in question and also the constraining power of the quantum conformality in question too. It will not constrain any more the detailed form of gravitational correlation functions of the quantum theory. The conformal symmetry will not be there and it will not protect the spectrum from the emergence of some spurious ghost states in it. This last thing will endanger the perturbative unitarity of the theory (and we do not speak here about the danger of unitarity breaking due to the HD nature of conformal gravity). Without the power of quantum  conformal symmetry, we may have unwanted states in the spectrum corresponding to the states from generic Stelle gravity, and not from the tree-level spectrum of conformal gravity, so we can see mismatch in counting number of degrees of freedom and also in their characters, namely whether they are spin-1 or spin-0, ghosts or healthy particles, etc.

Moreover, in pure conformal gravity described by the action simply $C^2$ without any supergravitational extension, we notice the somehow nomenclature problem with the presence of quantum conformality. Even barring the issue of conformal anomaly, the general pure gravitational theory has non-vanishing beta functions, so there is no UV-finiteness there. This implies that there is an RG running and scale-dependence of couplings and of various correlators on the renormalization energy scale. Hence already at the one-loop level one could say that scale-invariance is broken, which implies violation of conformal symmetry too. However, one can live with this semantic difference provided that there are no disastrous consequences of the conformal anomaly. One can adopt the point of view that the theory at the one-loop level is still good provided that the UV-divergent action is conformally invariant too, that is when one has only conformally invariant UV counterterms. (Although in the strict meaning having them implies non-vanishing beta function, RG running, loss of UV-finiteness and of scale-invariance.)
In our case the conformally invariant counterterms are only of the type $C^2$ and $\rm GB$, so if the $R^2$ counterterm is not present at the one-loop level, then we can speak about this preserved quantum conformality in the second sense. It happens this is exactly the situation we meet for quantum conformal gravity in $d=4$ at the one-loop level.

In order to see quantum conformality of one-loop level conformal gravity in $d=4$ described by the action, one first naively may try to take the limit $x\to+\infty$ from the expression for the $R^2$ sector of UV divergences from formula in (\ref{fourdergamma}). One would end up with the results, just a constant $A_0$, which is generally not zero. The whole story is again more subtle, since the limits in this case are again not continuous, although as we advocated above they are luckily also not divergent, when we want to send $\theta_R\to0$. In the end, we have only a finite discrepancy in numbers, which can be easily explained. As emphasized above in this case of the special reduced model we have the enhancement of symmetries and this new emergent conformal symmetry in the local version must be gauge-fixed too. This means that the kinetic operator needs to be modified and some new conformal gauge-fixing functional must be added to it for the consistency of the generalized Faddeev-Popov quantization prescription of theories with local gauge symmetries. This means that we will also have a new conformal gauge-fixing parameter (the fourth one), which can be suitably adjusted to provide again the minimality of the Hessian operator. Although, of course, the whole details of the covariant quantization procedure for conformal gravity are more delicate and more subtle, here we can just take a shortcut and pinpoint the main points of attention. When computing UV divergences using generally covariant methods like BV trace technology and functional traces of logarithms of operators, one also necessarily needs to add here the contribution of the third conformal ghosts, which are scalars from the point of view of Lorentz symmetry but they come with anti-commuting statistics. They are needed here because the conformal gravity theory is a natural HD theory and then third ghosts are necessary for covariant treatment of any gauge symmetry in the local form. It is true that for conformal local symmetry we do not need FP ghost fields (because of the reasons elucidated above), but we need a new third conformal ghost, which is moreover independent from the third ghost of diffeomorphism symmetry. Each symmetry with local realization comes with its own set of third ghosts, when the theory is with higher derivatives.
It is also well known that classically conformal fields (like massless gauge fields of electromagnetism and also of Yang-Mills theory) give contribution to divergences which is conformally invariant counterterm, so only of the type $C^2$ or $\rm GB$ terms. This can be understood easily as a kind of conformal version of the DeWitt-Utiyama argument used before. Hence, if the scalars of the anti-commuting type that we have to subtract were conformally coupled, then they will not contribute anything to the $R^2$ type of the counterterm. But we see from the formula in (\ref{fourdergamma}) that the $A_0$ coefficient there is non-zero, so only this one survives after the limit $x\to\infty$ is taken. To cancel the  $R^2$ counterterm is crucial for the hypothesized  conformal invariance of the conformal gravity also on the first loop quantum level. And this must be done by explicitly non-conformal fields with non-conformal contributions to divergences. They cannot be massless gauge fields, but they can be minimally, so not conformally coupled scalar fields. Here for the consistency of the whole formalism of the FP covariant quantization of conformal gravity, this role is played by the one real conformal third ghost with the kinetic operator $\square^2$.

The contribution of the third conformal ghosts is what we actually need to complete the whole process of the computation of the UV divergences in the conformal gravity model. We need them for the overall consistency since in this covariant framework we cannot a posteriori check the presence of all gauge invariances. Here we assume that on the first quantum loop level, the conformal gravity model enjoys fully diffeomorphism as well as conformal symmetry. The terms given in the covariant BV framework of computation all satisfy these requirements, so only we must be careful to take all these contributions into account. The contribution of the third conformal ghosts is like that of two real scalars coupled minimally (but not conformally to one's surprise) to the background gravitational metric, but of the ghost type. Indeed this means that we have to subtract the contribution of two scalars, which is of course, UV-divergent but after extracting the overall divergence there is only a finite number. This is the number that when subtracted now matches with number obtained after the naive limit $x\to\infty$ of the generic results from (\ref{eq: quadraticdivs}). We explain that we need to subtract two real scalars each one coming with the standard two-derivative GR-covariant box operator as the kinetic operator since in HD conformal gravity the operator between third conformal ghosts is of the $\square^2$ type as for the four-derivative theory. The limit to conformal gravity is discontinuous, but only in this sense that one has to take out also contribution of real scalar fields minimally coupled to gravitation. The first part of the limiting procedure, namely $x\to\infty$ is only a partial step and to complete the whole limiting procedure one must also deal properly with conformal symmetry. This applies not only to the coefficients in front of the $R^2$ term, where we see the mysterious but explainable $x$-dependence, but also to other coefficients in front of terms like $C^2$ and $\rm GB$ terms. Of course, for the last two terms the limits $x\to\infty$ do not change anything, but the contribution of third conformal ghosts makes impact and change the numerical results, which are luckily still finite in conformal gravity.
The coefficients in front of the $R^2$ and $\rm GB$ counterterms are also finite in generic four-derivative gravity (cf. with (\ref{fourdergamma})), however by these types of arguments with conformal gravity we cannot at present understand why the $x$-dependence happens only in front of the $R^2$ counterterm. Of course, in conformal gravity model, there is not  any $x$-dependence at all.

At the end, when one accounts for all these numerical contributions, one indeed finds that at the one-loop level in conformal gravity, the coefficient of the $R^2$ term is vanishing, so the quantum conformality is present in the second sense. And we have only conformally invariant counterterms in pure conformal gravity without any conformal matter, but there is still interesting RG flow of couplings there.
This also signifies that there is conformally invariant but non-trivial divergent part of the effective action with finite numerical constant coefficients, when the overall divergence is extracted. These finite coefficients arise in the two-step process. First as the limit $x\to\infty$ of a generic HD gravity and then by subtraction of UV-divergent contributions of two real scalars minimally coupled to gravitational field. Since these last contributions are known to be finite numbers multiplying the overall UV divergence, then this implies that the limit $x\to\infty$ of the generic expression in (\ref{eq: quadraticdivs}) must also give finite numbers. This explains why in perturbatively one-loop renormalizable model of conformal gravity in $d=4$ there are standard UV divergences, although this is a reduced model with $\theta_R=0$ and $N=0$ case. So the $x$-dependence in (\ref{fourdergamma}) must be as emphasized above that is with inverse powers of the fundamental ratio $x$ of the theory and in accord with what was schematically displayed in formula (\ref{eq: quadraticdivs}). Hopefully now the dissimilarities between the cases with $N=0$ and $N>0$ are more clear.

In short, we think that the only sensible reason, why we see completely different behaviour when going from $N=0$ to
$N=1$ class of theories is that the theory with $\omega_R=0$ and $N=1$ ceases to be conformally invariant in $d=4$. In a  different vein, the degeneracy of the kinetic operator $\hat H$ in the $\omega_C=0$ cases, for both $N=0$ and $N=1$, remains always the same.

This proves again and again that the case of conformal gravity is very special among all HD theories, in $d=4$ among all theories quadratic in gravitational curvatures. One can also study the phenomenological applications of the Weyl conformal gravity models to the evaporation process of black holes \cite{Bambi:2016yne,Bambi:2017ott} and also use the technology of RG flows (and also functional RG flows) in the quantum model of conformal gravity to derive some interesting consequences for the cosmology (like for example for the presence of dark components of the universe in \cite{Jizba:2019oaf,Jizba:2020hre,Rachwal:2021fhj}). Finally, the conformal symmetry realized fully on the classical level (and as we have seen also to the first loop level and perhaps also beyond) is instrumental in solving the issue with spacetime singularities \cite{Bambi:2016wdn,Modesto:2016max,Rachwal:2018gwu}, which are otherwise ubiquitous problems in any other model of generally covariant gravitational physics (both on the classical and quantum level). To resolve all singularities one must be sure that the conformality (Weyl symmetry) is present also on the full quantum level (and it is not anomalous there), so the resolution of singularities from the classical level (by some compensating conformal transformations) is not immediately destroyed by some dangerous non-conformal quantum fluctuations and corrections.

\subsection{More on limiting cases}
\label{s4.4}

 Here again we analyze closer the situation with various limits, when some coefficients in the gravitational action (\ref{eq: uvpart}) disappear. 

In a generic six-derivative theory, the trace of the logarithm  of the FP ghosts kinetic operator $\hat M$ and of the standard minimal $\hat C$ matrix are regular in
the limit $\omega_R\to0$, but not in the limit $\omega_C\to0$. For the $\hat C$ matrix this is understandable, because the $\gamma$ parameter
contains factor $\omega_C^{-1}$ in the minimal gauge. However, for the FP ghosts kinetic operator $\hat M$, this dependence was unexpected, because in the explicit definition of the  $\hat M$ operator there was never any singularity in $\omega_C$. Moreover, this singularity is even quadratic  in $\omega_C$ coefficient.

We also emphasize that in the general six-derivative theory the trace of the logarithm  of the Hessian operator  $\hat H$ is irregular both in the limits $\omega_R\to0$ and $\omega_C\to0$ separately. It seems
that in the total sum of contributions to the beta functions of the theory the singularity in $\omega_C$ cancels completely between ${\rm Tr}\ln\hat H$, ${\rm Tr}\ln\hat C$, and ${\rm Tr}\ln\hat M$, while the poles in $\omega_R$ remain and this is what is seen as a dependence of the final results on the non-negative powers of the fundamental ratio $x$.  For the definition of the $\hat H$ operator, if the limit $\omega_R\to0$ is taken, nothing bad is seen. This may be a partial surprise.  Of course, when the limit $\omega_C\to0$  is taken, then this operator does vanish on flat spacetime, so then its degeneracy is clearly visible.

The situation with limits ($\theta_R\to0$ or $\theta_C\to0$) in four-derivative theory we see as follows. The functional trace ${\rm Tr} \ln \hat C$  is regular in both limits. It actually does not depend on any gauge-fixing parameters here, despite that in its formal definition we used
the $\gamma$ parameter, which shows the $\frac{1}{\theta_C}$ singularity. The situation with the FP operator $\hat M$ is the same as previously, because the operator is identical as in the six-derivative theory case.
The operator $\hat H$ shows the problem with its definition only when the limit $\theta_C\to0$ is considered. The same is true for its trace of the functional  logarithm, which shows singularity in ${\theta_C}$ coupling coefficient up to the quadratic order. In this case and in the total sum of all contributions, we see only $\frac{1}{\theta_C}$ singularity to the
quadratic order. However, here the limit $\theta_R\to0$  is not continuous either, because the theory reaches a critical
point in the theory space with enhanced symmetry for $\theta_R=0$ (conformal enhancement of local symmetries) as it was explained in subsection \ref{s4.3}.

Let us also comment on what special happens in the computation of UV divergences for quadratic theory from the perspective of problems that we have initially encountered in six-derivative theory for the same computation. First, we established, in the middle steps of our computation for the results published in \cite{Rachwal:2021bgb}, that in the traces ${\rm Tr} \ln \hat H$ and ${\rm Tr} \ln \hat C$ in Stelle gravity there are no any dangerous $\frac{1}{y}=\frac{1}{2\omega_C-3\omega_R}$ poles (cf. \cite{Avramidi:1985ki}). The cancellations happen separately within each trace. Second thing is that we found that the trace ${\rm Tr} \ln \hat C$ surprisingly completely does not depend  on the gauge-fixing parameter $\gamma$, which was needed and used in the initial definition of the  $\hat C$ operator in (\ref{weight}). Finally, one can notice that the addition of the Gauss-Bonnet term in $d=4$ spacetime dimension, does not change anything for $R^2$, $R_{\mu\nu}^2$, and $R_{\mu\nu\!\rho\sigma}^2$ divergences (as it was expected), because its variation is a topological term in $d=4$.

In this last part, we use the schematic notation for various gravitational theories, when we do not write, for simplicity, the coupling coefficients in front of various terms since they are not the most important for the considerations here.
In the case of six-derivative theories, it is impossible to obtain the results for the cases with $\omega_C=0$ or $\omega_R=0$ by any limiting procedures of the corresponding results obtained for the general six-derivative theory with $\omega_R\neq0$ and  $\omega_C\neq0$.
These reduced theories have different bilinear parts, with degenerate forms of the kinetic operator and our calculation methods break down here. Similarly, one can calculate the beta functions in a theory with $R^2$ only and this was done indirectly many times. One can also calculate UV divergences in $C \square C + R^2$ theory or in an analogous $R \square R + C^2$ theory, but this is actually not easy to do. But then we cannot easily extract these results from our general calculation done  in $ C \square C + R \square R$ six-derivative theory. The simple reason is that all these theories have different amount and characteristics of degrees of freedom and the transition from
one to another at quantum level is complicated (and to some extent unknown).

Moreover, we remark, that the results for beta functions in models $C \square C+ R^2$ (or $C \square C+ R^2 + C^2$) or in $R \square R + C^2$ (or $R \square R + C^2+ R^2$) could be obtained by though different computations than what we have done here. We summarize that the six-derivative gravitational theory to be renormalizable must contain both terms of the type $C\square C$ and $R\square R$. Then the kinetic operator (Hessian) between gravitational fluctuations and the graviton's propagator are well-defined. In all other models, there is not a balance between the number of derivatives in the vertices of the theory and in all gauge-invariant pieces of the propagator, so the theory behaves badly regarding perturbative UV divergences at higher loops. This does not mean that the computation of UV divergences at one-loop level is forbidden, just only that usually these are not all divergences in the theory, they  may not be the UV-leading ones anymore or the theory does not have decent control over all of them.

For the strictly non-renormalizable theory with the leading in the UV term $C\square C$ we can have additions of  various subleading terms which do not change the fact of non-renormalizability. We can add terms (separately or in conjunction) of the following types: $\omega_\Lambda$ (cosmological constant term), $R$ (E-H term), $R^2$ (Starobinsky's term), $C^2$ (Weyl square term). The UV-leading part of the Hessian still is defined as it contains six-derivative differential operator understood on flat spacetime and between tensorial fluctuations, so derived from the terms quadratic in curvatures. The Hessian is non-degenerate. (It has to be non-degenerate here, because the GR-covariant box operator is here only a spectator, and the Hessian must be ``almost'' non-degenerate for the case of conformal gravity with the action $C^2$.) The flat spacetime propagator can be defined only if we make addition of $\omega_\Lambda$, or $R$ or $R^2$  terms -- this is because of the problematic part of it proportional to the projector $P^{(0)}$ which must for the consistency of the inverting procedure for the whole propagator be non-zero. This scalar part (spin-0 part) is sourced from any scalar term or from the cosmological constant term. If only the $C^2$ term is added, then the propagator still is ill-defined. Still these additions do not change the fact that the theory is non-renormalizable, if there is not an accompanying six-derivative term of the form $R\square R$. As for the final results for UV divergences in these extended models, naively one would think that there are no additional UV divergences proportional to terms with four derivatives of the metric (namely to terms $R^2$, $C^2$ and $\rm GB$), because of the limit $\omega_R\to0$ and the dependence on the $x$ ratio in (\ref{sixdergamma}) in the linear way. We would naively think that divergences with $R^2$ and $\rm GB$ terms are the same as in (\ref{sixdergamma}). The only problematic one could be this proportional to the $C^2$ term since the limit gives already divergent results (so ``doubly'' divergent) -- this would mean that the coefficient of the  $C^2$ divergence is further divergent and renormalization of just $C^2$ does not absorb everything at the one-loop level. Since the model is non-renormalizable we cannot trust this computation and these limits at the end, especially if they give divergent results. But this probably means that we cannot sensibly define the $C^2$ counterterm needed for the UV renormalization in these theories. In a sense an attempt of adding $\omega_\Lambda$ or $R$, or $R^2$  terms to regularize the theory $C\square C+C^2$, or even the simplest one, just $C\square C$, is unsuccessful so we perhaps still cannot trust there in the final results just given by two $B_0$ coefficients of UV divergences proportional to terms $R^2$ and $\rm GB$, while the $C^2$ divergences are never well-defined in this class of models.

Instead, in the case of the reduced model with $R \square R$ UV-leading action, one may keep some hope
that the results for the $C^2$ counterterms will be finite at the end, but maybe still discontinuous,
despite the non-renormalizability of the model with $R \square R$ action (plus possible lower derivative
additions to regularize it as it was mentioned above). Maybe in this reduced models the results of the
projection procedure of the UV-divergent functional of the effective action of the theory onto the sector
with only $C^2$ terms will result here in giving sense to pure $C^2$ divergences in this limiting model.
(Here we may try to resort to some projection procedure for the functional with UV divergences since in these non-renormalizable models, one may expect to find more divergences than just of the form of  $C^2$ and $R^2$ as presented initially in  (\ref{sdiv}). There could exist new UV divergences, which contain even more than four derivatives on the background metric tensor, even in $d=4$ case.)
But the final finite value may be discontinuous and may not be obtainable by the naive limit $x\to0$ of the
expression for the divergent term in the $C^2$ sector of UV divergences, so it may not be just $B_0$
there. This remark about possible discontinuities may apply also to coefficients in front of divergent
terms of the type $R^2$ and ${\rm GB}$. They may still end up with some finite definite values for this
model, but probably they are not the same as the coefficients $B_0$ of these terms from (\ref{eq:
sexticdivs}), so we probably will be able to see here another discontinuities in  taking the naive limit
$x\to0$.

These above results about discontinuities and negative consequences due to the overall
non-renormalizability of the two considered above reduced models, are also enforced by the analysis of
power counting of UV divergences. One can try to perform the ``worst case scenario'' analysis of one-loop
integrals and the results show complete lack of control over perturbative UV divergences in such reduced
models. This is even worse that in the case of off-shell E-H gravity considered in $d=4$ dimensions,
which is known to be one-loop off-shell perturbatively non-renormalizable theory. In the latter case the
superficial divergence of the divergence $\Delta$ is bounded at the one-loop level ($L=1$) in formula
(\ref{pc2}) by the value $4$. In general, at the arbitrary loop order we have the formula for power
counting reads then
\begin{equation}
\Delta+d_\partial=4+2(L-1),
\end{equation}
which if we concentrate on logarithmic UV divergences only (with $\Delta=0$), we get that at the one-loop
level for all Green functions we need counterterms with up to $d_\partial=4$ partial derivatives on the
metric tensor. At the two-loop level we instead need to absorb the divergence term with $d_\partial=6$
partial derivatives as this was famously derived by Goroff and Sagnotti
\cite{Goroff:1985sz,Goroff:1985th}. The counterterm that they have found was of the form of the $C^3$
GR-covariant term and its perturbative coefficient at the two-loop order does not vanish, and this
implies that the whole UV-divergent term does not vanish even in the on-shell situation. But still we
know which counterterms to expect at the given loop order and the absorption of UV divergences works for
all divergent Green functions of the QG model.

The situation in the reduced models of the type $C\square C$ or $R\square R$ in the leading UV terms is
much worse even at the one-loop level from naive power counting there. One sees that different
GR-covariant counterterms are needed to absorb divergences in different divergent Green functions of the
quantum model at the one-loop level, so the counting does not stop at the two-point function level. We
think that despite these tremendous difficulties, one still can compute the divergent parts of the
effective action and the actual computations are very tedious and still possible. This is provided that
one can invert the propagator, so one has some non-vanishing parts in both gauge-invariant parts of it
with the spin-0 and spin-2 projectors. So, it is at present practically impossible to do computation in
the pure models $C\square C$  or $R \square R$ only. We know that they give contributions in momentum
space proportional to $k^6$ in the spin-2 and spin-0 parts of the propagator respectively, while other
parts are not touched. In order to regularize the theory and to give sense to the perturbative propagator
around flat spacetime, one has to add the regulator terms as this was mentioned above. Let us assume that
they give contributions to the other sector of the spin projectors in the graviton's propagator of the
form $k^{-m}$, where $m$ is some integer and $m<6$, they may likely also give additional subleading
contributions to the main respective part of the propagator which was there with six derivatives in the
UV regime correspondingly to the spin-2 sector in $C\square C$ theory and to the spin-0 sector in
$R\square R$ model. The values of $m$ are respectively: $m=0$ for the cosmological constant addition (it
still regularizes the propagator but very, very weakly), $m=2$ when E-H term is added (it contributes
both to the spin-0 and spin-2 parts), $m=4$ when either $R^2$ or $C^2$ terms are added (they contribute
exclusively in their respective sectors).

Since now after the regularization of the graviton's propagator, its behaviour is still very unbalanced
in the UV regime between different components, then one sees the following results of the analysis of UV
divergences at the one-loop order. First, the general gravitational vertex is still with six derivatives,
while the propagator is $k^{-6}$ in the best (most suppressed) behaviour and $k^{-m}$ is the worst
behaviour in the other components. For the most dangerous situation we have to assume that the overall
behaviour of the propagator is in the worst case, so with the UV scaling of the form $k^{-m}$.
Then the relation between the number of derivatives in a general gravitational vertex and in the
propagator is broken and this is a reason for very bad behaviour with UV divergences here. Such relation
is  typically present even in non-renormalizable models, like in E-H gravity. The lack of this relation
means that now the result for $d_\partial$ of any Feynman graph $G$ depends on the number of external
graviton lines $n_g$ emanating from the one-loop diagram. Previously in the analysis of power counting
there was never any dependence on this $n_g$ parameter. This is a source for problems even bigger when
one increases $n_g$.  For definiteness we can assume that $n_g>1$ since here we will not be interested in
vacuum or tadpole diagrams and quantum corrections to them. Now, for a general diagram $G$ with $n_g$
external graviton lines, the worst situation from the point of view of UV divergences is for the
following topology of the diagram, namely there is one loop of gravitons (so called ``ring of gravitons'')
in the middle with $n_g$ 3-leg vertices joined by $n_g$ propagators. In the case when we concentrate on
logarithmic divergences $D=0$, we get the following results for the quantity $d_\partial$ which tells us
how many derivatives we must have in the corresponding counterterm to absorb the divergence,
\begin{equation}
d_\partial=4+n_g(6-m)
\label{pc3}
\end{equation}
for the graph contributing one-loop quantum corrections to the $n_g$-point Green function. One sees that
this $d_\partial$ grows without a bound even at the one-loop level, when $n_g$ grows, so in principle to
renormalize the theory at the one-loop level one would need already infinitely many GR-covariant terms,
if one does not bound the number of external legs of Green functions that must be considered here. A few
words about explanation of numbers appearing in the formula (\ref{pc3}). The $4$ there is the number of
spacetime dimensions (integration over all momenta components at the one-loop level), while the $(6-m)$
factor comes from the difference between the highest number of derivatives in the vertex, i.e. $6$
compensated by the worst behaviour in some propagator components given in the UV by $k^{-m}$ only.

Moreover, there are precisely $n_g$ segments of the structure propagator joined with 3-leg vertex to
create a big loop. This behaviour signals complete lack of control over perturbative one-loop divergences
even at the one-loop level. Moreover, they have to be absorbed in the schematic terms of the type
\begin{equation}
\nabla^{4+4n_g-n_gm+2i}{\cal R}^{n_g-i},
\end{equation}
for the index $i$ running over integer values in the range $i=0,1,2,\ldots n_g-2$, where we only
mentioned the total number of covariant derivatives not specifying how they act on these general
gravitational curvatures. This is because this is an expression for the quantum dressed one-loop Green
function with $n_g$ graviton legs on the flat spacetime, so terms with more curvatures than $n_g$ will
not contribute to absorb these divergence of flat Green $n_g$-point function. We only mentioned here the
really the worst situation, where the divergence may be finally absorbable not only by the highest
curvature terms of the type $\nabla^{4+4n_g-n_gm}{\cal R}^{n_g}$ but also for terms with smaller number
of curvatures (up to ${\cal R}^2$ terms and in the precise form ${\cal R} \square^{\frac12 n_g(6-m)}
{\cal R}$). We neglect writing counterterms here which are total derivatives and which are of the
cosmological constant type. These are then the needed counterterms off-shell at the one-loop level in
such general reduced model.

To make it more concrete, we will analyze the cases of $m=0,2$ and $4$ with special attention here in
these badly non-renormalizable models for some small numbers of legs of quantum dressed Green functions.
We have that at $m=0$ to absorb UV divergences from the $2$-point function we need generic counterterms of
the form: ${\cal R}\square^j{\cal R}$ with the exponent $j$ running over values $j=0,1,2,3,4,5,6$, while
to renormalize a three-point function one needs previous terms and possibly new terms of the type
$\nabla^j{\cal R}^3$ with $j=0,\ldots,16$ and for four-point functions new terms are of the type
$\nabla^j{\cal R}^4$ with $j=0,\ldots,20$, etc. for higher Green functions (for $n_g$-leg correlators one
needs $j$ up to $j_{\rm max}=4n_g+4$). When we regularize by adding the E-H term with $m=2$ the situation
is slightly better, but then to absorb UV divergences from the $2$-point function we need generic
counterterms of the form: ${\cal R}\square^j{\cal R}$ with the exponent $j$ running over values
$j=0,1,2,3,4$, while to renormalize a three-point function one needs previous terms and possibly new
terms of the type $\nabla^j{\cal R}^3$ with $j=0,\ldots,10$ and for four-point functions new terms are of
the type $\nabla^j{\cal R}^4$ with $j=0,\ldots,12$, etc. for higher Green functions (for $n_g$-leg
correlators one needs $j$ up to $j_{\rm max}=2n_g+4$). Finally, we can add terms of the type $R^2$ and
$C^2$ for the regularization purposes. In such final case to be considered here one has that to absorb UV
divergences from the $2$-point function we need generic counterterms of the form: ${\cal R}\square^j{\cal
R}$ with the exponent $j$ running over values $j=0,1,2$, while to renormalize a three-point function one
needs previous terms and possibly new terms of the type $\nabla^j{\cal R}^3$ with $j=0,\ldots,4$ and for
four-point functions new terms are of the type $\nabla^j{\cal R}^4$ with $j=0,\ldots,4$, etc. for higher
Green functions (for $n_g$-leg correlators one needs $j$ up to $j_{\rm max}=4$ here independently on the
number of legs $n_g$). Still in this last case one sees that one needs infinitely many counterterms to
renormalize the theory at the one-loop level, although the index $j$ of added covariant derivative is
bounded by the values $4$, still one needs more terms with more powers of gravitational curvatures $\cal R$.

This shows how badly non-renormalizable are these models already at the one-loop level and that any
control over perturbative UV divergence is likely lost, when the number of external legs is not bounded
here from above. These reduced models are examples of theories when the dimensionality and the number of
derivatives one can extract from the vertices and propagators of the theory differ very much. Previously
in quantum gravity models these two numbers were identical which leads to good properties of control
over perturbative divergences (renormalizability, super-renormalizability and even UV-finiteness). With
these reduced models we are on the other bad extreme of  vast possibilities of QG models. But seeing them
explicitly proves to us how precious is the renormalizability property and why we strongly need them in
HD models of QG, in particular how we need super-renormalizability in six-derivative QG models.

The arguments above convince us to think that there is no hope to get convergent results for the front
coefficient coming with the $C^2$  counterterm in the UV-divergent part of the effective action in the
considered here model with the only presence of the $C\square C$ as the leading one in the UV. This may
signify that here there exists another UV-divergent term (perhaps of the structure like $C\square^n C$),
which contains more derivatives, and this could be a reason why the coefficient in front of the $C^2$
term is itself a divergent one. The presence of such new needed counterterms with more derivatives can be
motivated by the analysis of power counting of UV divergences in this reduced unbalanced model, which is
also presented below. Even if the higher $C \square^n C$ type of UV divergence is properly extracted and
taken care of, then we can still be unable to properly define and see as convergent the divergence
proportional to the four-derivative term $C^2$. Even such a projection of the UV-divergent functional of
the theory onto the sector with only $C^2$ terms will not help here in giving sense to pure $C^2$
divergences in this limiting model. However, this remark does not need necessarily to apply to
coefficients in front of divergent terms of the type $R^2$ and ${\rm GB}$. They may still end up with
some finite definite values for this model, but probably they are not the same as the coefficients $B_0$
of these terms from (\ref{eq: sexticdivs}), so we could be able to see here another discontinuity in
taking the limit $x\to+\infty$.

On the other hand, for the strictly non-renormalizable theory with the leading in the UV term $R\square R$ we can have additions of similar  various subleading terms which do not change the fact of non-renormalizability. We can add terms (separately or in conjunction) of the following types: $\omega_\Lambda$, $R$, $R^2$, or $C^2$. The UV-leading part of the Hessian still is not well-defined as it should contain six-derivative differential operator understood on flat spacetime and between tensorial fluctuations, while from the term $R\square R$ we get only operator between traces of metric fluctuations $h=\eta^{\mu\nu}h_{\mu\nu}$ (between spin-0 parts), so derived from the terms quadratic in curvatures present in the UV regime. Probably the degeneracy of the Hessian operator can be easily lifted out, if we add one of the $\omega_\Lambda$, or $R$ or $C^2$  terms. If only the $R^2$ term is added, then the Hessian still is degenerate. Similarly, the flat spacetime propagator can be defined only if we make addition of $\omega_\Lambda$, or $R$ or $C^2$  terms -- this is because of the problematic part of it proportional to the projector $P^{(2)}$ which must for the consistency of the inverting procedure for the whole propagator be non-zero. This tensorial part (spin-2 part) is sourced exclusively from any GR-invariant term built out with Weyl tensors in adopted here Weyl basis of terms or from the E-H term or from the cosmological constant term. If only the $R^2$ term is added, then the propagator still is ill-defined. Still these additions do not change the fact that the theory is formally non-renormalizable, if there is not an accompanying six-derivative term of the form $C\square C$.

As for the final results for UV divergences in these extended models, naively one would think that there are no new UV divergences proportional to terms with four derivatives of the metric (namely to terms $R^2$, $C^2$ and $\rm GB$), because of the limit $\omega_C\to0$ and the dependence on the $x$ ratio in (\ref{sixdergamma}) in the linear way. We would naively think that divergences with $R^2$ and $\rm GB$ terms are the same as in (\ref{sixdergamma}). The only problematic one could be this proportional to the $C^2$ term since the limit gives already  constant result, namely the $B_0$ coefficient. Since the model is non-renormalizable we cannot trust this computation and these limits at the end, even if they give here convergent results. But this probably means that we cannot sensibly define the $C^2$ counterterm needed for the renormalization of these theories. When we include additions to the action, which remove the degeneracy of the flat spacetime graviton's propagator, then at least the perturbative computation using Feynman diagrams can be attempted in such a theory. Although of course, in this case the different parts of the propagator have different UV scalings, so the situation for one-loop integrals is a bit unbalanced and there is not a stable control over perturbative UV divergences, when for example one goes to the higher loop orders. Probably new counterterms (with even higher number of derivatives) will be at need here. Making additions of some subleading terms from the point of view of the UV regime, may help in defining the unbalanced perturbative propagator, but still one expects (due to energy dimension considerations) that these additions do not at all influence the quantitative form of UV divergences with four-derivative terms, so these ones which are leading in the UV. These additions are needed here only quantitatively and on the formal level to let the computation being done for example using Feynman diagrams with some mathematically existing expressions for the graviton's propagator. In a sense adding $\omega_\Lambda$ or $R$, or $C^2$  terms regularizes the theory $R\square R+R^2$ or even the simplest one, just $R\square R$, so we can perhaps trust there in the final results just given by all three $B_0$ coefficients of UV divergences proportional to terms $R^2$, $C^2$ and $\rm GB$. This can be motivated by the observation that here the limits $\omega_C\to0$ or $\theta_C\to0$ respectively do not enhance any symmetry of the model in question, so they can be naively and safely taken. But we agree that this case requires a special detailed and careful computation to prove this conjectural behaviours.

In the general case of badly non-renormalizable theories, with both $\omega_C=0$ or $\omega_R=0$, one trusts more the computation using Feynman diagrams and around flat spacetime than of the fully GR-covariant BV method of computation. For the former one only needs to be able to define properly the propagator -- all physical sectors of it, and for this purpose one can regularize the theory by adding the term $\omega_\kappa R$, which is a dynamical one with the smallest number of derivatives and for which flat spacetime is an on-shell vacuum background. (In this way, we exclude adding the cosmological constant term $\omega_\Lambda$, which would require adding some source and the flat spacetime propagator could not be considered anymore in vacuum there.) Then in such regulated non-renormalizable theory one can get results around flat spacetime and in Fourier momentum space, and then at the end one can take the limit $\omega_\kappa\to0$.  The results for some UV divergences in these non-renormalizable models must be viewed as projected since higher-derivative (like 6-derivative and even higher) infinities may be present as well. These last results must coincide with the ones we obtained in (\ref{sixdergamma}), when the proper limits of $\omega_C\to0$  or $\omega_R\to0$  are taken. We notice that adding the E-H term, which is always a good regulator, changes the dynamics very insignificantly for these higher-derivative models and the results from Feynman diagram computations can be always derived. Instead for the leading in the UV regime part of the Hessian operator, which is a crucial element for the BV method of computation, addition of just $\omega_\kappa R$ does not help too much and the operator is still degenerate since it is required that all its sectors are with six-derivative differential operators: are non-vanishing and non-degenerate there.

We propose the following procedure for the derivation of correct limiting cases analyzed here. First the theory with $\omega_C\neq0$, $\omega_R\neq0$ and $\omega_\kappa\neq0$ is analyzed using Feynman diagram approach. The results for UV divergences must be identical to the ones found in (\ref{sixdergamma}) using the BV technique. They do not show any singularity at this moment. In Feynman diagram computation one can take the limit $\omega_C\to$ or $\omega_R\to0$, while the propagator and perturbative vertices are still well-defined. In these circumstances we have that still $\omega_\kappa\neq0$. We admit that the theory loses now its renormalizability properties, but we just want to project the UV-divergent action onto the terms with the structure of three GR-covariant terms $C^2$, $R^2$  and $\rm GB$. For this the method of Feynman graph computation is still suitable since it only requires the well-defined propagator, but it can work even in badly non-renormalizable theories. This is like taking the naive limits of $\omega_C\to$ or $\omega_R\to0$ in the results from (\ref{sixdergamma}) respectively, regardless which way they were obtained. One justifies the step of taking these limits by recalling that we are still working in the Feynman diagram approach, and not with the BV technique. Finally, one sends $\omega_\kappa\to0$ hoping that this does not produce any finite discontinuity in the results for four-derivative UV divergences. This is justified by dimensional analysis arguments provided also earlier in this article and by the fact that except in the case of conformal gravity theory in $d=4$, there is no any enhancement of local symmetries in the limit $\omega_\kappa\to0$. Then one gets the sense for the limits considered in these sections.

This analysis concludes the part with special limiting cases of extended six-derivative theories, where one of the coefficients $\omega_R$ or $\omega_C$ is to be set to zero. Probably the same considerations of some limits can be repeated very similarly (with the exception of the conformal gravity case) for the Stelle quadratic gravity models, but we will skip this analysis here since it can be found in the literature.


\section{Stability of HD theories}
\label{s5}

Above we have seen that HD theories of gravitation are inevitable due to quantum considerations. They also come with a lot of benefits that we have discussed at length before like super-renormalizability and the possibility for UV-finiteness. However, it is also well known that they have their own drawbacks and problems. One of the most crucial one is the issue of unitarity of the scattering ($S$-matrix) in perturbative framework. This is of course in the situation when we can discuss the scattering problems, so when we can define asymptotic states in interacting gravitational background, so when the gravitational spacetime is asymptotically flat. In more generality, the related issue is of quantum stability of the theory.

In general, in literature about general HD theories there exist various proposals for solutions of these perennial problems. We can mention here a few of them: $PT$-symmetric quantum theory, Anselmi-Piva fakeon prescription, non-local HD theories, benign ghosts as proposed by Smilga, etc. Below we will try to describe some of their methods and show that the problems with unitarity or with the stability of the quantum theory can be successfully solved. We also provide arguments thanks to Mannheim \cite{Mannheim:2016lnx,Mannheim:2014ypa} that the gravitational coupled theory must be without problems of this type, if the original matter theory was completely consistent.

We first express the view that the stability of the quantum theory is fundamental, while the classical theory may emerge from it only in some properly defined limits. Hence we should care more about the full even non-linear stability on the quantum level and some instabilities on the classical level may be just artifacts of using classical theory which cannot be defined by itself without any reference to the original fundamental quantum theory. An attempt to understand the stability entirely in classical terms may be doomed to clearly fail since forgetting about the quantum origin may be here detrimental for the limiting process. If the quantum theory is stable and unitarity is preserved, then this is the only thing we should require since the world is in its nature quantum and physically we know that it is true that $\hbar=1$ in proper units, rather than $\hbar\to0$, so the classical limit may be only some kind of illusion. If there are problems with classical stability analysis like this done originally by Ostrogradsky, then this may only mean that the classical theory obtained this way neglects some important features that were relevant on the quantum level for the full quantum stability of the system.

First, in Anselmi-Piva prescription one solves completely the unitarity issue for HD theories by invoking fakeon prescription to take properly into account the contribution of particles which in the spectrum are related to higher derivatives theories and which typically are considered as dangerous for the unitarity of the theory. The presence of a
particle with negative residue called a ghost at the classical level makes the theory not unitary in its
original quantization based on the standard Feynman prescription \cite{Stelle:1976gc} of encircling the poles for the loop integrals.
A new quantum prescription, as recently introduced by Anselmi and
Piva \cite{Anselmi:2017lia,Anselmi:2017yux,Anselmi:2018kgz} was based on the earlier works by
Cutkosky, Landshoff, Olive, and Polkinghorne \cite{Cutkosky:1969fq}. The former authors invented a
procedure for the Lee-Wick theories \cite{Lee:1969fy,Lee:1970iw}, which allow them to
tame the effects typically associated to the presence of ghosts in the Stelle's theory.
In this picture,
the ghost problem (also known as unitarity problem) is solved consequently at any perturbative order in the loop
expansion \cite{Anselmi:2018kgz} done for the loop integrals which need to be computed in any QFT, if one requires higher order accuracy.

At the classical level, the ghost particle (or
what Anselmi and Piva define as ``fakeon'', because this particle understood as a quantum state can
only appear as a virtual particle and inside perturbative loops) is removed from the perturbative spectrum of the theory.
This is done by
solving the classical equations of motion for the fakeon field by the mean of a very specific combination of
advanced plus retarded Green's functions and by fixing to zero the
homogeneous solution of resulting field equations
\cite{Anselmi:2018bra,Anselmi:2018tmf}. This is then
equivalent to removing the complex ghosts in the quantum theory
from the spectrum of asymptotic quantum states by hand. However, this choice
and this removal decision
is fully preserved and protected by quantum corrections, hence it does not
invalidate the unitarity of the $S$-matrix at higher loop orders.

Such prescription of how to treat virtual particles arising due to HD nature of the theories
is very general and can be applied to both real
and complex ghosts, and also to normal particles, if one wishes to. (Every particle can be made fake,
 so without observable effects on the unitarity of the theory.) In particular, this prescription
is crucial in order to make perturbatively unitary the theory
proposed by Modesto and Shapiro in
\cite{Modesto:2015ozb,Modesto:2016ofr} which comes under the name of
``Lee-Wick quantum gravity''. The latter class of theories is based
on the general gravitational higher-derivative actions as proposed by Asorey, Lopez,
and Shapiro \cite{Asorey:1996hz}. In this range of theories,  we can safely state to
have a class of super-renormalizable or UV-finite and unitary
higher-derivative theories of QG. In order to
guarantee tree-level unitarity, the theory in
\cite{Modesto:2015ozb,Modesto:2016ofr} has been
constructed in such a way that it shows up only complex conjugate poles in the graviton's propagator,
besides the standard spin-2 pole typically associated with the normal massless graviton particle with two polarizations.
 Afterwards, the new prescription by Anselmi
and Piva \cite{Anselmi:2018kgz} guarantees the  unitarity of this theory at any perturbative
order in the loop expansion.

We also emphasize that the Stelle's quadratic theory in gravitational curvatures \cite{Stelle:1976gc} with the
Anselmi-Piva prescription is the only strictly renormalizable theory
of gravity in $d=4$ spacetime dimensions, while the theories proposed in
\cite{Modesto:2015ozb,Modesto:2016ofr} are from a large (in principle infinite)
class of super-renormalizable or UV-finite models for quantum gravity.

Next, in the other approach pioneered by Bender and Mannheim to higher-derivative theories and to non-symmetric and non-Hermitian quantum mechanics \cite{Bender:2007wu,Bender:2008gh}, one exploits the power of non-
Hermitian $PT$-symmetric quantum gravity. Here, the basic idea is that the gravitational Hamiltonian in such theories (if it can be well-defined), is not a Hermitian operator on the properly defined Hilbert space of quantum states, rather it is only $PT$-symmetric Hamiltonian.
Then some eigenstates of such a Hamiltonian may correspond to non-stationary solutions of the original classical wave equations. They would indeed correspond in the standard classical treatment to the Ostrogradsky instabilities. The famous example are cosmological  run-away solutions or asymptotically non-flat gravitational potentials for the black hole solutions. The problem of ghosts manifests itself  already on the classical level of equations of motion, where one studies the linear perturbations and its evolution in time. For unstable theories, the perturbations growth is without a bound in time. But in some special solutions, like for example present in models of conformal gravity, these instabilities are clearly avoided and then one can speak that ghosts are benign in opposition to them being malign in destroying the unitarity of the theory.
Such benign ghosts \cite{Smilga:2004cy,Smilga:2013vba} are then innocent for the issues of perturbative stability.

In the $PT$-symmetric approach to HD theories at the beginning, one cannot determine the Hilbert space by looking at the $c$-number propagators of quantum fields. In this case, one has to from the start quantize the theory and construct from the scratch the Hilbert space, which is different than the original naive construction based on the extension of the one used normally for example for two-derivative QFT's. With this new Hilbert space and with the non-Hermitian (but $PT$-symmetric) Hamiltonian the theory revealed to be quantum-mechanically stable. This is dictated by the construction of the new Hilbert space and the structure of the Hamiltonian operator.
In that case the procedure of taking the classical limit, results in the definition of the theory in one of the Stokes wedges and in such a region the Hamiltonian is not real-definite and the corresponding classical Hamiltonian is not a Hermitian operator. Therefore, the whole discussion of Ostrogradsky analysis is correct as far as the theory with real functions and real-valued Hamiltonians is concerned, but it is not correct for the theory which corresponds to the quantum theory which was earlier proven to be stable quantum-mechanically. The whole issue is transmitted and now there is not any problem with unitarity or classical stability of the theory, but one has to be very careful in attempts to define the classical limiting theory.

We also repeat here arguments proposed by Mannheim about stability of the resulting gravitation-matter coupled theory \cite{Mannheim:2016lnx,Mannheim:2014ypa}. First we take some matter two-derivative model (like for example standard model of particle physics, where we have various scalars, fermions and spin-1 gauge bosons). This theory as considered on flat Minkowski background is well known to be unitary so it gives $S$-matrix of interactions with these properties. The model can be said that it is also stable on the quantum level. Now, we want to couple it to gravity, or in other words put it on gravitational spacetime with non-trivial background in such a way that the mutual interactions between gravitational sector and matter sector are consistent. This, in particular, implies that the phenomena of back-reaction of matter species on geometry are not to be neglected. The crucial assumption here is that this procedure of coupling to gravity is well behaved and for example, it will not destroy the unitarity properties present in the matter sector. We know that the theory in the matter sector is stable and also its coupling to geometry should be stable on the full quantum level. After all, this is just simple coupling procedure (could be minimal coupling) to provide mutual consistent interactions with the background configurations of the gravitational field.

Next, on the quantum level described, for example, by functional path integral, we can decide to completely integrate out matter species still staying on the general gravitational background. As emphasized in section \ref{s1}, such procedure in $d=4$ spacetime dimensions generate effective quantum gravitational dynamics of background fields with higher derivatives, precisely in this case there are terms of the type $C^2$ and $R^2$ (the latter term is absent when the matter theory is classically conformally invariant). In other words, the resulting functional of the quantum partition function of the total coupled model is a functional of only background gravitational fields. This last reduced or ``effective'' functional is given by the functional integral over quantum fluctuations of gravitational field of the theory given classically by the action with these HD above terms. Let us recall now what we have done, namely we have simply integrated out all quantum matter fields, which is an identity transformation for the functional integral representation of the partition function $Z$ of the quantum coupled theory. Since this transformation does not change anything, then also the resulting theory of gravitational background must necessarily possess the same features as the original coupled theory we started with. Since the first theory was unitary, then also the last one theory of pure gravity but with higher-derivative terms must be unitary too. We emphasized that both theories give the same numerical values of the partition function $Z$ understood here as the functional of the background spacetime metric. In the first theory the integration variables under functional integrals are quantum matter fields, while in the second case we are dealing with pure gravity so we need to integrate over quantum fluctuations of the gravitational fields. In the last case the model, which gives the integrand of the functional integral is given by  the classical action $S_{\rm HD}$, so it contains necessarily higher derivatives of the gravitational metric field.

There also exist possibilities that ghosts or classical instabilities one sees on the classical level thanks to Ostrogradsky analysis disappear. This may happen if for example, some very specific (or fine-tuned) initial or boundary conditions are used for solving non-linear higher-derivative classical equations of motion of the theory. It is not excluded as proven by Smilga that some instabilities may  go away if one analyzes such special situations.

Various cures have been proposed in the literature for dealing with the ghosts-tachyon issue: Lee-Wick prescription
\cite{Lee:1969fy,Lee:1970iw}, fakeons \cite{Anselmi:2018kgz,Anselmi:2018bra,Anselmi:2019nie,Anselmi:2018tmf}, non-perturbative numerical methods \cite{Tkach:2012an,Smilga:2013vba,Smilga:2004cy,Kaku:1982xt,Tomboulis:1977jk,Tomboulis:1980bs,Tomboulis:1983sw}, ghost instabilities \cite{Salles:2014rua,Peter:2017xxf,deOSalles:2018eon}, non-
Hermitian $PT$-symmetric quantum gravity based on $PT$-symmetric quantum mechanics \cite{Bender:2007wu,Bender:2008gh}, etc (see also \cite{Christodoulou:2018jbn,Briscese:2018oyx,Donoghue:2019fcb,Asorey:2018wot,Briscese:2018bny,Briscese:2019rii}). One might even entertain the
idea that unitarity in quantum gravity is not a fundamental concept. So far, there is no a consensus in the community which solutions is the correct one. The unfortunate prevalent viewpoint is that none of the
proposed solutions solves conclusively and completely the problem. And it seems that sadly the solutions proposed in the literature are not compatible and are unrelated to each other.

Therefore all the arguments given above should convince the reader that the HD (gravitational) theories are stable on the full quantum level. In particular, this means that for situations in which we can define asymptotic states (like for asymptotically flat spacetimes) the scattering matrix between fluctuations of the gravitational field is unitary on the quantum level and both perturbatively and non-perturbatively.

\section{Conclusions}
\label{s6}

In this contribution, we have discussed the HD gravitational theories, in particular six-derivative gravitational theories. First, we motivated them by emphasizing their various advantages as for the models of consistent Quantum Gravities. We showed that six-derivative theories are even better behaved on the quantum level than just four-derivative theories, although the latter ones are very useful regarding scale- and conformal invariance of gravitational models. Moreover, the models with four-derivative actions serve as good starting examples of HD theories and they are reference points for consideration of six- and higher order gravitational theories. We first tried to explain the dependence of the beta functions in six-derivative theories by drawing analogies exactly to these prototype theories of Stelle gravity. We also emphasize that only in six-derivative gravitational models we have the very nice features of super-renormalizability and the narrow but still viable option for complete UV-finiteness. This is why we think super-renormalizable six-derivative theories have better control over perturbative UV divergences and give us a good model of QG, where this last issue with perturbative divergences is finally fully under our control and theoretical understanding.

In the main part of this paper, we analyzed the structure of perturbative one-loop beta functions in six-derivative gravity for couplings in front of terms containing precisely four-derivative in the UV-divergent part of the effective actions. These terms can be considered as scale-invariant term since couplings in front of them are all dimensionless in $d=4$ spacetime dimensions. Our calculation for these divergences was done originally in the Euclidean signature using the so-called Barvinsky-Vilkovisky trace technology. However, the results are the same also in the Minkowskian signature independently which prescription one uses to  rotate back to the physical relativistic Lorentz signature case, whether this is standard Wick rotation, or the one using Anselmi-Piva prescription using fakeons. This is because they are the leading divergences in the UV regime, and hence they do not completely depend how the rotation procedure is done from Euclidean to Minkowskian and how for example the contributions of arcs on the complex plane is taken into account since the last ones give subleading contributions to the UV-divergent integrals. Moreover, the calculations of beta functions that we presented in this paper has very nice and important features of being renormalization scheme-independent since they are done at the one-loop, but the expressions we get for them are valid universally. These are exact beta functions since they do not receive any perturbative corrections at the higher loop orders since the six-derivative gravitational
theory is super-renormalizable in $d=4$.  Another part of good properties of the beta functions obtained here are the complete gauge independence and also independence on the gauge-fixing parameters one can use in the definition of the gauge-fixing functional. These last two properties are very important since in general gravitational theory we have the access to perturbative computation only after introducing some spurious element to the formalism which are related to gauge freedoms (in this case these are diffeomorphism symmetries). We modify the original theory (from the canonical formalism) by adding various additional fields and various spurious nonphysical (gauge) polarizations of mediating gauge bosons (in our case of gravitons) in order also to preserve relativistic invariances. These are redundancies that have to be eliminated when at the end one wants to compute some physical observables.  Therefore, it is  very reassuring that our final results are completely insensitive to these gauge-driven modifications of original theories.

Our beta functions being exact and with a lot of nice other properties, constitute one significant part of the accessible observables in the QG model with six-derivative actions. Their computation is a nice theoretical exercise, which of course from the sense of algebraic and analytic methods used in mathematical physics has its own sake of interest. However, as we emphasized above these final results for the beta functions may have also meaning as true physical observables in the model of six-derivative QG theories.

We described in greater detail the analysis of the structure of beta functions in this model. First we used arguments of energy dimensionality and the dependence of couplings on the dimensionless fundamental ratio of the theory $x$. Next, we tried to draw a comparison between the structure of 4-derivative gravitational Stelle theory and six-derivative theory in $d=4$ dimensions. We showed the dependence on the parameters $x$ is quite opposite in two cases. The case with four-derivative theory is exceptional because the model without any $R$ term in the action (and also without the cosmological constant term) enjoys enhanced symmetry and then the quantum conformal gravity is renormalizable at the one-loop order, so then it is a special case of a sensible quantum physical theory (up to conformal anomaly problems discussed earlier). We also remark that in the cases of $x\to0$ and $x\to\infty$ the generic six-derivative theories are badly non-renormalizable. This was the source of the problem with attempts to obtain sensible answers in these two limits. Non-renormalizability problem must show in some place in the middle or at the end of the computation to warn us that at the end we cannot trust in the final results for the beta functions in these cases. In these two cases this problem showed indeed in two different places and the logical consequences of this were strongly constraining the possible form of the rational $x$-dependence of these results. Thanks to these considerations we were able finally to understand whether the positive or inverse powers of the ratio $x$ must appear in the final results for beta functions in question.
Of course, we admit that this analysis is a posteriori since we first derived the results for the divergences and only later tried to understand the reasons behind these results. But eventually we were able to find a satisfactory explanation.

And there are a few of additional spin-offs of the presented argumentation. First, we can make predictions about the structure of beta functions in 8-derivative and also of other higher-derivative gravitational theories with the number of derivatives in the action which is bigger than 4 and 6 (analyzed in this paper). We conjecture that the structure should be very similar to what we have seen already in the generic case with six derivatives, so only positive powers of the corresponding fundamental ratio $x$ of the theory, and probably only in the sector with $C^2$ type of UV divergences. Another good side effect is that we provide first (to our knowledge) theoretical explanation of the structure of beta functions as seen in four-derivative case of Stelle theory in $d=4$ spacetime dimensions. It is not only that the theory with $C^2$ action is exceptional in $d=4$ dimensions; we also ``explained'' these differences based on an extension of the theory to include higher-derivative terms like with 6-derivative and quantify to which level the theory with $C^2$ action is special and how this reflects on the structure of its one-loop beta functions. We remark that in Stelle gravity (or even in its subcase model with conformal symmetry based on the $C^2$ action), there are contributions to beta functions originating from higher perturbative loops since the super-renormalizability argument based on power counting analysis does not apply here. Our partial explanation of the structure of the one-loop beta functions in Stelle theory in $d=4$ uses a general philosophy that to ``explain'' some numerical results in theoretical physics, one perhaps has to generalize the original setup and in this new extended framework looks for simplifying principles, which by reduction to some special cases show explicitly how special are these cases not only qualitatively but also quantitatively and what this reduction procedure implies on the numbers one gets as the results of the reduction.
For example, one typically extend the original framework from $d=4$ spacetime dimension fixed condition to more general situation with arbitrary $d$ and then draw the general conclusion as a function of $d$ based on some general simple principles. Then finally, the case of $d=4$ is recovered as a particular value one gets when the function is evaluated for $d=4$. And this should explain its speciality.
In our case, we extended the four-derivative theory by adding terms with six derivatives and in this way we were able to study a more generic situation. This was in order to understand and explain the structure of divergences in the special reduced case of conformal gravity in $d=4$ and of still generic four-dimensional Stelle theory. We think that this is a good theoretical explanation which sheds some light on the so far mysterious issue of the structure of beta functions. One can also see this as another advantage of why it is worth to study generalizations of higher-derivative gravitational actions to include terms with even more higher number of derivatives, like 6-derivative, 8-derivative actions, etc.

Finally, here we can comment on the issue of experimental bounds on the values of the ratio $x$. Since it appears in six-derivative gravitational theory the constraints on its possible values are very weak. Slightly stronger constraints apply now for the corresponding value of the ratio in four-derivative Stelle gravitational theory in $d=4$ case. Since the main reason for higher-derivative modifications of gravity comes because of consistency of the coupled quantum theory, then one would expect that the stringent bounds would come from experimental measurement in the real domain of true quantum gravity. Of course, right now this is very, very far, if possible at all, future for experimental gravitational physics. This is all due to smallness of gravitational couplings characterized by $G_N$ proportional inversely to the Planck mass $M_P\sim10^{19}\,{\rm GeV}$. And in the quantum domain of elementary particle physics this scale is bigger than any energy scale of interactions between elementary quanta of matter. This implies that also quantum gravitational interactions are very weak in strength. Hence the only experimental/observational bounds we have on the coefficients in front of higher-derivative terms come from the classical/astrophysical domain of gravitational physics and they are still very weak. To probe the values of the coefficients in front of six-derivative terms, one would have to really perform a gravitational experiment to the increased level of accuracy between elementary particles in the full quantum domain, which is now completely unfeasible. Hence, we have to be satisfied with already existing very weak bounds, but this lets us to freely consider theoretical generic situation with arbitrary values of the ratio $x$ since maybe only (very far) future experiments can force us theoreticians to consider some more restricted subset or interval for the values of the $x$ ratio as consistent with observed situation in the Nature. For the moment it is reasonable to consider  and explore theoretically all possible range of values for the $x$ ratio and also both possible signs. (Only the case with $x=0$ is excluded as a non-renormalizable theory that we have discussed before.)

In the last section of this contribution, we commented on the important issue of stability of higher-derivative theories. We touched both the classical and quantum levels, while the former should not be understood as a standalone level on which we can initially (before supposed quantization) define the classical theory of the relativistic gravitational field. We followed the philosophy that the quantum theory is more fundamental and it is a starting point to consider various limits, if it is properly quantized (in a sense that the quantum partition function is consistently defined, regardless of how we get there to its form, no matter which formal quantization procedure we have been following).
One of such possible limit is the classical limit where the field expectation values are large compared to characteristic values as found in the microworld of elementary particles. And also the occupation number for bosonic states are large number (of the order of Avogadro number for example). Then we could speak about coherent states which could define well classical limit of the theory. Such procedure has to be followed in order to define HD classical gravitational theory. We emphasized that quantum theory is the basis and classical theory is the derived concept, not vice versa. On the quantum level we shortly discussed various approaches present in the literature to solve the problems with unwanted ghost-like particle states. They were classified in two groups: theories with $PT$-symmetric Hamiltonian and theories with Anselmi-Piva prescription instead of the Feynman prescription to take into account contributions of the poles of the ghost but without spoiling the unitarity issue. On the quantum level, we considered mainly the issue with unitarity of the scattering matrix since this seems the most problematic one. The violation of unitarity would signal the problem with conservation of the probability of quantum processes. Something that we cannot allow to happen in quantum-mechanical framework for the isolated quantum system (non-interacting with the noisy decohering and dissipative or some thermal environment). Of course, such an analysis was tailor-made for the cases of gravitational backgrounds on which we can define properly the scattering process.

In general, the scattering processes are not everything we can talk about in quantum field theories even for on-shell quantities. The analysis of some on-shell dressed Green functions may also show some problems with quantum stability of the system. Therefore, we briefly also described the results of the stability analysis, both on the classical and quantum level and to the various loop accuracy in QG models. This analysis is in principle applicable to the case of any gravitational background, more general than the one coming with the requirement of asymptotic flatness. We also mentioned that in some cases of classical field theory the analysis of classical exact solutions shows that the very special and tuned solutions are without classical instabilities and they are well-defined for any time starting with very special initial or boundary conditions. For example, here we can mention the case of so-called benign ghosts of higher-derivative gravitational theories as proposed by Smilga some time ago. This should prove to the reader that we are dealing with the theories which besides a very interesting structure of perturbative beta functions, are also amenable to solve the stability and unitarity issues in these theories, both on the quantum as well as on the classical level. With some special care we can exert control and HD gravitational theories are stable quantum-mechanically and this is what matters fundamentally.

\begin{acknowledgements}
 We would like to thank I. L. Shapiro, L. Modesto and A. Pinzul for initial comments and encouragement about this work. L. R. thanks the Department of Physics of the Federal University of Juiz de Fora for kind hospitality and FAPEMIG for a technical support. Finally, we would like to express our gratitude to the organizers of the ``Algebraic and analytic methods in physics'' -- AAMP XVIII conference for accepting our talk proposal and for creating a stimulating environment for scientific online discussions.
\end{acknowledgements}

\bibliographystyle{actapoly}
\bibliography{biblio}

\end{document}